\documentclass[12pt,epsf,revtex]{article}
\usepackage{amsmath,amssymb}
\usepackage{graphicx}
\begin{document}

\def\a{\alpha}
\def\b{\beta}
\def\c{\varepsilon}
\def\d{\delta}
\def\e{\epsilon}
\def\f{\phi}
\def\g{\gamma}
\def\h{\theta}
\def\k{\kappa}
\def\l{\lambda}
\def\m{\mu}
\def\n{\nu}
\def\p{\psi}
\def\q{\partial}
\def\r{\rho}
\def\s{\sigma}
\def\t{\tau}
\def\u{\upsilon}
\def\v{\varphi}
\def\w{\omega}
\def\x{\xi}
\def\y{\eta}
\def\z{\zeta}
\def\D{\Delta}
\def\G{\Gamma}
\def\H{\Theta}
\def\L{\Lambda}
\def\F{\Phi}
\def\P{\Psi}
\def\S{\Sigma}

\def\o{\over}
\def\beq{\begin{eqnarray}}
\def\eeq{\end{eqnarray}}
\newcommand{\gsim}{ \mathop{}_{\textstyle \sim}^{\textstyle >} }
\newcommand{\lsim}{ \mathop{}_{\textstyle \sim}^{\textstyle <} }
\newcommand{\vev}[1]{ \left\langle {#1} \right\rangle }
\newcommand{\bra}[1]{ \langle {#1} | }
\newcommand{\ket}[1]{ | {#1} \rangle }
\newcommand{\EV}{ {\rm eV} }
\newcommand{\KEV}{ {\rm keV} }
\newcommand{\MEV}{ {\rm MeV} }
\newcommand{\GEV}{ {\rm GeV} }
\newcommand{\TEV}{ {\rm TeV} }
\def\diag{\mathop{\rm diag}\nolimits}
\def\Spin{\mathop{\rm Spin}}
\def\SO{\mathop{\rm SO}}
\def\O{\mathop{\rm O}}
\def\SU{\mathop{\rm SU}}
\def\U{\mathop{\rm U}}
\def\Sp{\mathop{\rm Sp}}
\def\SL{\mathop{\rm SL}}
\def\tr{\mathop{\rm tr}}

\def\IJMP{Int.~J.~Mod.~Phys. }
\def\MPL{Mod.~Phys.~Lett. }
\def\NP{Nucl.~Phys. }
\def\PL{Phys.~Lett. }
\def\PR{Phys.~Rev. }
\def\PRL{Phys.~Rev.~Lett. }
\def\PTP{Prog.~Theor.~Phys. }
\def\ZP{Z.~Phys. }


\baselineskip 0.7cm


\begin{titlepage}

\begin{flushright}
IPMU12-0076\\
ICRR-report-614-2012-3\\
\end{flushright}

\vskip 1.35cm
\begin{center}
{\large \bf The Lightest Higgs Boson Mass in the MSSM\\ with Strongly Interacting Spectators
}
\vskip 1.2cm
Jason L. Evans$^1$, Masahiro Ibe$^{1,2}$ and Tsutomu T. Yanagida$^1$
\vskip 0.4cm
$^1${\it Kavli IPMU, TODIAS, University of Tokyo, Kashiwa 277-8583, Japan}\\
$^2${\it ICRR, University of Tokyo, Kashiwa 277-8582, Japan}
\vskip 1.5cm

\abstract{
We propose a new mechanism for producing a Higgs boson mass near $125$\,GeV within the MSSM. By coupling the MSSM Higgs boson to a set of strongly interacting fields, large corrections to the Higgs quartic coupling are induced. Although the Higgs doublets do not participate in the strong dynamics, they feel the effects of the strongly coupled sector via (semi-)perturbative interactions. These same strong dynamics are also capable of generating the $\mu$-term.
Additionally, this strong sector is in the conformal window, which drives the couplings to an infrared
fixed point and naturally generates model parameters of the appropriate size.
}
\end{center}
\end{titlepage}

\setcounter{page}{2}

\section{Introduction}
A Higgs boson mass of about 125 GeV indicated by the ATLAS and CMS collaborations\,\cite{ATLAS, CMS} strongly supports a supersymmetric (SUSY) extension of the standard model\,\cite{Fayet}. However, its mass is slightly larger than expected\,\cite{Okada:1990vk}
in the minimal supersymmetric Standard Model (MSSM). Thus,
various mechanisms to enhance the Higgs boson mass
have subsequently been proposed.

Within the MSSM, the simplest approaches require either a very large SUSY breaking scale, i.e.
a gravitino mass of $m_{3/2}=10-100$\,TeV\,\cite{Higgsmass,Wells:2004di,Giudice:2011cg,Hall:2011jd,pure}
or $A$-terms  of order a few TeV
(for recent discussions\,\cite{Evans:2011bea, Heinemeyer:2011aa,Arbey:2011ab,Draper:2011aa}).
These are consistent scenarios, but they create tension with naturalness in the MSSM.
This is because heavy stops and large $A$-terms induce large radiative corrections to the Higgs potential necessitating a fine tuning of the MSSM parameters in order to realize electroweak symmetry breaking at the weak scale.

On the other hand, if additional contributions to the Higgs self quartic coupling
are present, a $125$\,GeV Higgs boson mass is possible even for a relatively low SUSY-breaking scale and a small  $A$ term.
Under these conditions, the tension with naturalness becomes much more mild. There have been, in fact, two classes of models proposed that generate additional contributions to the Higgs self quartic coupling. One method is to introduce a gauge singlet that couples to the Higgs
doublets\,\cite{NMSSM,Ellis:1988er,Ellwanger:2011aa}
and the other is to add an additional $U(1)$ gauge
interaction\,\cite{Batra:2003nj,Maloney:2004rc,Endo:2011gy}.

In this paper, we propose an alternative mechanism
for generating a relatively large Higgs quartic coupling in the MSSM.
Our mechanism assumes the existence of an additional strongly coupled sector whose influence is communicated to the MSSM via the Higgs doublets.
Although the Higgs doublets couple with this strong sector, they are not charged under the new strong gauge group. However, they feel the effects of this strong sector via
(semi-)perturbative interactions. These semi-perturbative interactions
to the strongly interacting spectators
induce a large Higgs quartic coupling, which in turn enhances the Higgs boson mass.

We also show that the $\mu$-term can be dynamically generated
by coupling the Higgs boson to strongly interacting spectators.
Interestingly, we find that an extension of the strongly interacting sector
which is consistent with grand unification is in the conformal window.
In this extension, the model parameters appropriate for a Higgs mass of $125$\,GeV
are naturally obtained by renormalization group running to the infrared (IR) fixed point.

The organization of the paper is as follows.
In section\,\ref{sec:SSD}, we discuss the generic structure of the strongly interacting sector which enhances the effective quartic coupling of the Higgs doublet.
In section\,\ref{sec:CSSD}, we construct a model which can also explain the origin of the $\mu$-term.
The refinements in section\,\ref{sec:conformal} push our model into the conformal window and are consistent with grand unification.
There we show that the appropriate parameter values are provided by a fixed point
of the renormalization group equations.

\section{Strongly Interacting Spectators}\label{sec:SSD}
\begin{figure}[tb]
\begin{center}
  \includegraphics[width=.7\linewidth]{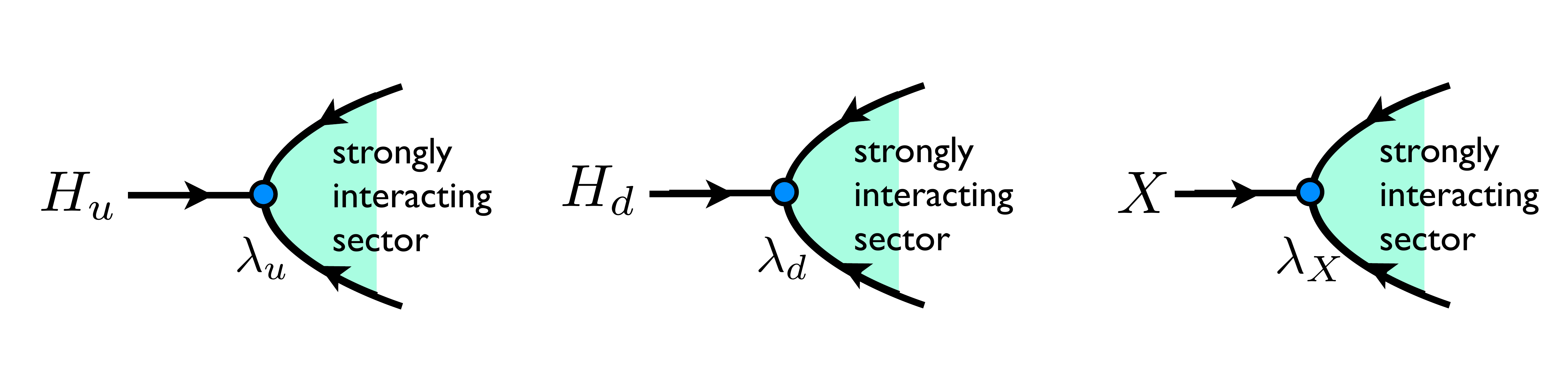}
\caption{\sl \small
An illustrative picture of the couplings to the ``Strongly  interacting spectators".
The Higgs doublets and the SUSY breaking spurion couple
to a strongly interacting sector via $\lambda_{u,d,X}$, respectively.
}
\label{fig:SSD}
\end{center}
\end{figure}
Before discussing explicit models,
let us summarize some generic features of our model with strongly coupled spectators.
In our model, we assume that the Higgs doublets $H_u$ and $H_d$
are elementary superfields but couple to a strongly interacting sector
in the superpotential with coupling constants $\lambda_u$ and $\lambda_d$, respectively.
That is, the elementary Higgs doublets couple to some operators consisting of strongly
interacting fields
\begin{eqnarray}
 W_H = \lambda_u  H_u {\cal O}_u + \lambda_d  H_d {\cal O}_d \ .
\end{eqnarray}
Here, the Higgs doublets are elementary superfields and have the usual MSSM gauge and Yukawa interactions. This simple structure is one of the advantage of our model.
In other models of this type that can realize a heavier lightest Higgs boson, the Higgs doublets (and top quark) are composite fields
(Refs.\,\cite{Harnik:2003rs,Fukushima:2010pm,Csaki:2012fh}).

In addition, we also assume that the spectators couple
to a SUSY breaking spurion field $X  =  M_X +  F_X\h^2 $ via a coupling constant $\lambda_X$;
\begin{eqnarray}
 W  = \lambda_X X {\cal O}_X \ .
\end{eqnarray}
In the following discussion, we include the effects of the supersymmetric expectation value of $X$, $M_X$,
as well as the supersymmetry breaking expectation value $F_X$, and treat each of these as fixed parameters.%
\footnote{
The dynamical generation of the expectation values of $X$
can be done by scaling down the cascade SUSY breaking mechanisms\,\cite{Ibe:2010jb}.
}
As we will show below, these mass parameters, $M_X$ and $\sqrt{F_X}$,  are required to be of $O(1)$\,TeV.

Finally, for the models discussed below,
some of the fields in the strongly interacting sector become massless in the limit of $\lambda_X \to 0$.
In such models, to exclude regions with tachyonic masses we require
\begin{eqnarray}
\label{eq:NonTac}
 |\lambda_X M_X|^4 \gtrsim     |\lambda_X F_X|^2   \ ,
\end{eqnarray}
where the uncertainty in the inequality represents the incalculable effects of the strongly interacting sector.

In Fig.\,\ref{fig:SSD}, we show an illustrative picture of the interactions
between the strongly coupled spectators, and the elementary Higgs doublets
and the SUSY breaking spurion field.
The strongly interacting sector is assumed to confine at a dynamical scale $\Lambda_H$.
Below this scale the strongly interacting sector decouples from the MSSM sector.
A concrete model is developed in the next section.

\subsection{Effective Higgs Quartic Coupling}
\begin{figure}[tb]
\begin{center}
  \includegraphics[width=.7\linewidth]{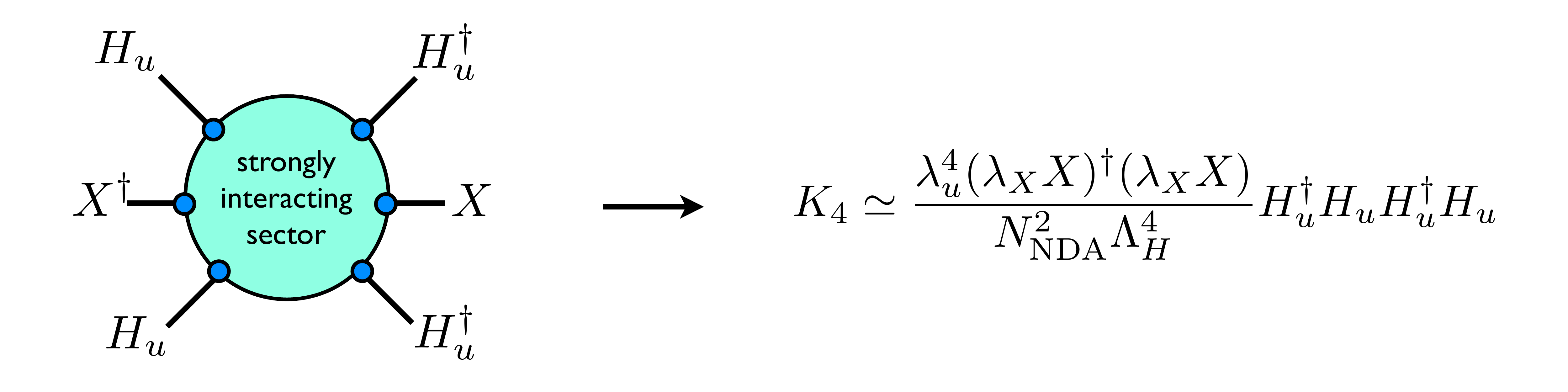}
\caption{\sl \small
An illustrative diagram for the effective quartic term in the K\"ahler potential.
}
\label{fig:K4}
\end{center}
\end{figure}

With the above construction, we immediately find that there are contributions to the quartic
couplings of the Higgs scalar potential from the strongly interacting spectator fields. These spectator fields generate an effective K\"ahler potential of,
\begin{eqnarray}
\label{eq:K4}
 K_{4} \simeq  \frac{\l_u^4 (\l_X X)^\dagger (\l_X X)  }{N_{\rm NDA}^2\L_H^4}
{H}_u^\dagger {H}_u {H}_u^\dagger {H}_u \ ,
\end{eqnarray}
(see Fig.\,\ref{fig:K4}).
The coefficient $N_{NDA}$ is expected to be $N_{NDA}\sim 4\pi$
by naive dimensional analysis\,\cite{Luty:1997fk,Cohen:1997rt}.
We also obtain terms involving $H_d$'s via similar diagrams.
In the above expressions, we have not shown the gauge superfields which should
be inserted between $H_u^\dagger$ and $H_u$.%
\footnote{
The effects of these higher dimensional operators on the MSSM Higgs bosons
have been discussed extensively in Refs\,\cite{Dine:2007xi,Carena:2009gx}.
}

For the above effective K\"ahler potential, the Higgs potential has an additional
contribution,
\begin{eqnarray}
\label{eq:V4}
 V &\simeq & \frac{\lambda_{\rm eff}}{4} | h^\dagger h|^2 \ , \cr
 \lambda_{\rm eff} &\simeq&  \frac{4\l_u^4}{N_{\rm NDA}^2}
 \frac{M_H^4}{\L_H^4}\frac{x^2}{\l_X^2} \sin^4\b\ ,
\end{eqnarray}
where we have introduced the notation $x = F_X/M_X^2$ and $M_H = \lambda_X M_X$.
Using the Higgs mixing angle $\b$, we have replaced $H_u$ by the light Higgs boson $h$.
The quartic terms in the K\"ahler potential involving $H_d$ will also contribute to the $\lambda_{\rm eff}$ but will be suppressed by $\cos\b$, $\mu/M_H$ or $\mu/\Lambda_H$.
In the following discussion, we have omitted these contributions by assuming $\tan\b$
is rather large, i.e. $\tan\b\gtrsim 5$,
and the $\mu$-term is much smaller than $M_H$ and $\L_H$.
The following arguments, however, can be extended to include these contributions in a straightforward way.%
\footnote{
If one includes the $H_d$ contributions to the effective Higgs potential, there could be an enhancement of $h\to \gamma\gamma$ \cite{Blum:2012kn}.
}

In Eq.\,(\ref{eq:V4}), we have assumed that the coupling constants $\lambda_{u,d,X}$ as well as $M_X$ and $F_X$ are real.
These assumptions can be validated by field redefinitions in the models discussed in the
following sections.
The sign of the effective quartic coupling $\lambda_{\rm eff}$, on the other hand, cannot be determined
due to the incalculability of the strongly interacting sector.
As we will see, however, it is possible to show that the effective quartic coupling
is positive valued for the perturbative limit of the strongly coupled sector.
Armed with the results of our perturbative examples,
we assume that the effective quartic coupling is positive valued even in the strongly
coupling limit.

\begin{figure}[tb]
\begin{minipage}{.49\linewidth}
\begin{center}
  \includegraphics[width=.9\linewidth]{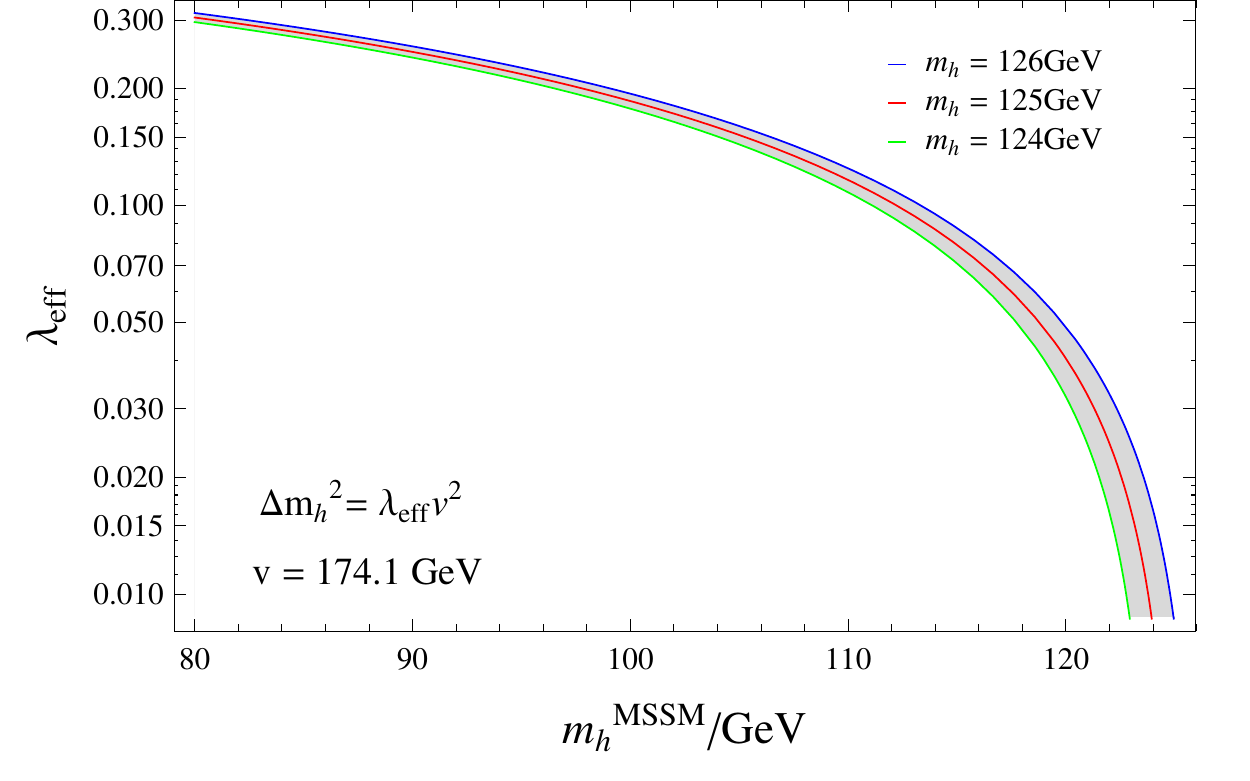}
  \end{center}
  \end{minipage}
 \begin{minipage}{.49\linewidth}
 \begin{center}
  \includegraphics[width=.7\linewidth]{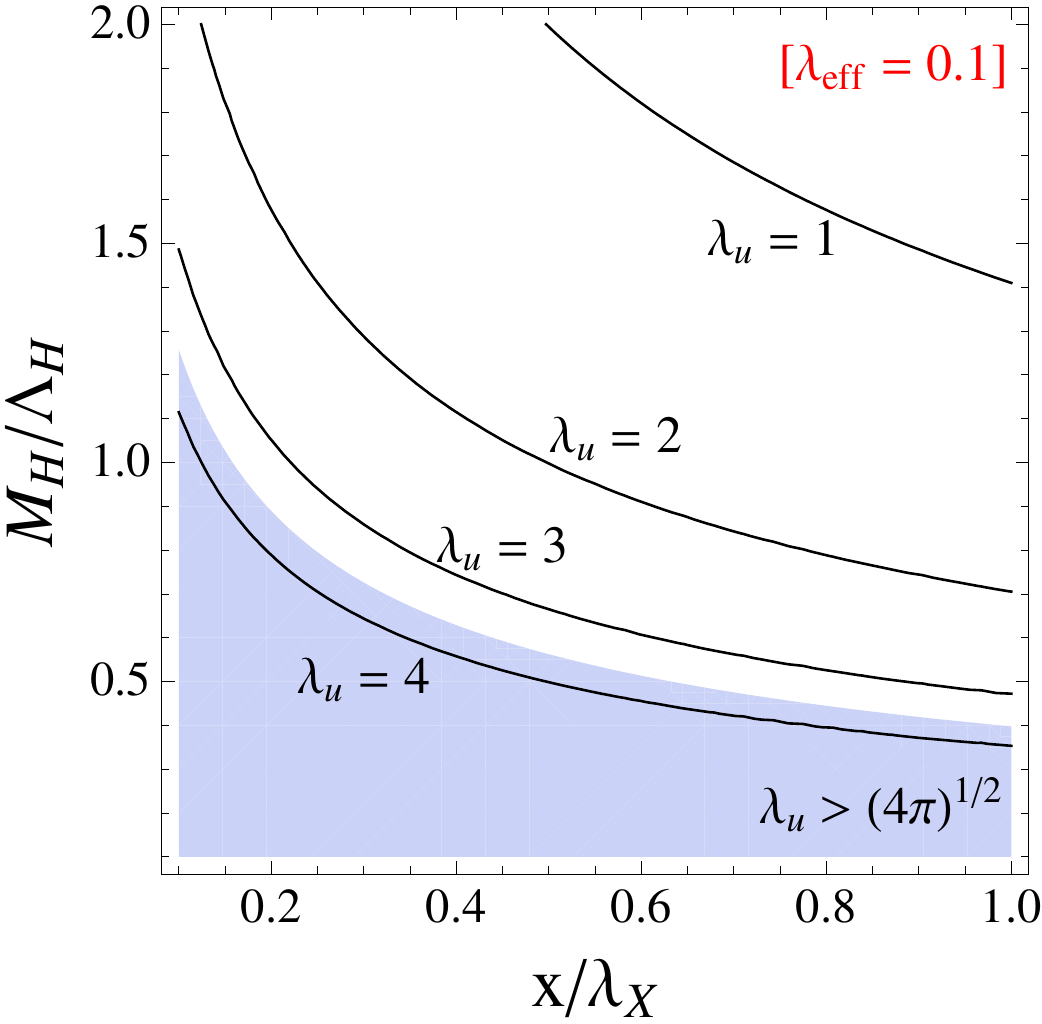}
   \end{center}
  \end{minipage}
\caption{\sl \small
Left) The required quartic coupling for $m_h = 124-126$\,GeV as a function of
$m_{h}^{\rm MSSM}$.
For example, $m_h = 125$\,GeV requires $\lambda_{\rm eff} \simeq 0.08$
if the MSSM contribution gives $m_{h}^{\rm MSSM}=115$\,GeV.
Right) Contours of the coupling constant $\l_u$
which realizes $\l_{\rm eff}=0.1$ as a function of $(x/\l_X, M_H/\L_H)$.
We have taken $\sin\b\simeq 1$.
In the shaded region, the coupling constant $\l_u$ also becomes rather
strong and the perturbative treatment of $\l_u$ is less reliable.
}
\label{fig:lambda}
\end{figure}

Including the effective quartic term leads to
an additional contribution to the lightest Higgs boson mass in the MSSM,
\begin{eqnarray}
 m_h^2 &=& m_h^{{\rm MSSM}\,2} +  {\mit \D} m_h^2   \ ,\cr
 {\mit \D} m_h^2 &=& \lambda_{\rm eff} v^2\ ,
 \end{eqnarray}
 where $ m_h^{{\rm MSSM}} $ denotes the lightest Higgs boson mass in the MSSM
 and   $v\simeq 174.1$\,GeV.%
 \footnote{Here, $m_h^{\rm MSSM}$ means the radiatively corrected Higgs boson mass in the MSSM.}
 In Fig.\,\ref{fig:lambda}, we have shown the required values of the effective quartic term necessary
 to realize a lightest Higgs
boson mass of $m_h = 124-126$\,GeV
as a function of the MSSM contribution $ m_h^{{\rm MSSM}}$.
The figure shows, for example, that a lightest Higgs boson mass of
$m_h = 125$\,GeV requires $\lambda_{\rm eff} \simeq 0.08$
for
$m_{h}^{\rm MSSM}=115$\,GeV.

Such an effective quartic coupling can be realized
for $\lambda_u= O(1)$ with $M_H \simeq \Lambda_H$ and $\lambda_X \simeq x$,
\begin{eqnarray}
 \lambda_{\rm eff}= 0.025 \times \lambda_u^4 \left(\frac{4\pi}{N_{\rm NDA}}\right)^2
 \frac{M_H^4}{\Lambda_H^4} \frac{x^2}{\lambda_X^2}\ ,
\end{eqnarray}
(see Eq.\,(\ref{eq:V4})).
In the right panel of Fig.\,\ref{fig:lambda}, we show contours of $\l_u$ which realize
$\l_{\rm eff}= 0.1$.
This figure clearly shows that the quartic coupling $\l_{\rm eff} = O(0.1)$ can be realized
for $\l_u= O(1)$,  $x/\l_X =O(1)$ and $M_H \sim \L_H$.%
\footnote{For $M_H \gg \L_H$, the expansion of the effective K\"ahler potential in Eq.\,(\ref{eq:K4}) is
no longer valid.}

\subsection{Soft Masses from Strongly Interacting Spectators}
\begin{figure}[tb]
\begin{center}
  \includegraphics[width=.7\linewidth]{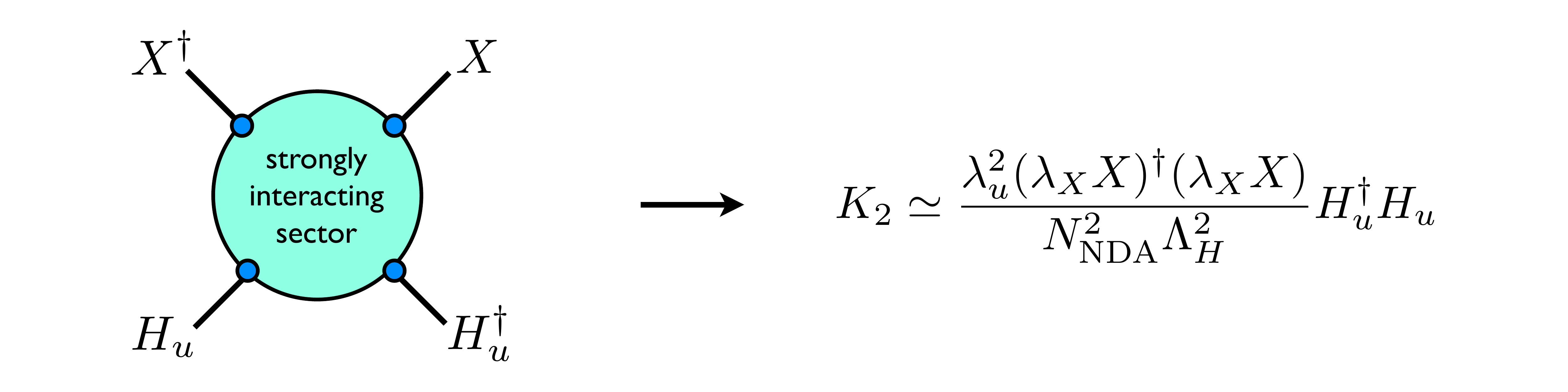}
\caption{\sl \small
An illustrative diagram for the effective quadratic term in the K\"ahler potential.
}
\label{fig:K2}
\end{center}
\end{figure}
Before closing this section,  let us discuss the soft squared masses
of the Higgs doublet which are generated through $\lambda_{u,d}$, its coupling to the strongly coupled sector.
Along with the effective quartic term in Eq.\,(\ref{eq:V4}),
the strong dynamics also generate effective SUSY breaking mass terms
via the effective K\"ahler potential in Fig.\,\ref{fig:K2};
\begin{eqnarray}
\label{eq:K2}
 K_{2} \simeq  \frac{\l_u^2 (\l_X X)^\dagger (\l_X X)  }{N_{\rm NDA}^2\L_H^2}
{H}_u^\dagger {H}_u \ .
\end{eqnarray}
This term leads to an additional contribution to the soft mass squared  of $H_u$,
\begin{eqnarray}
\label{eq:mHu}
{ \mit \D} m_{H_u}^2 \simeq
  \frac{\l_u^2}{N_{\rm NDA}^2}\frac{M_H^2}{\L_H^2}\frac{x^2}{\lambda_X^2} M_H^2
  \simeq \frac{\lambda_{\rm eff}}{4\lambda_u^2} \Lambda_H^2\ .
\end{eqnarray}
Therefore, the contribution to the Higgs soft mass squared from the spectator dynamics
can be less than $O(1)$\,TeV as long as the dynamical scale, $\Lambda_H$ is of $O(1)$\,TeV.

\section{$\mu$-term from Strongly Interacting Spectators}\label{sec:CSSD}
In the above sections, we have discussed the effective quartic term generated by the
strongly interacting sector examining the generic features of the spectator fields.
In this section, we discuss an ambitious extension of these ideas that also generates
the supersymmetric Higgs mixing term, i.e. the $\mu$-term.

\subsection{Confining of Spectators by Strong Dynamics}\label{sec:deformed}
\begin{table}[t]
\caption{\sl \small
The strongly interacting sector is charged under $SU(3)_H$ which is based on supersymmetric
QCD with three-flavors.
We embed the Standard Model gauge groups $SU(2)_L\times U(1)_Y$ into
the subgroups of the maximal global symmetry $U(3)\times U(3)$.
}
\begin{center}
\begin{tabular}{|c|c|c|c|c|}
\hline
&$SU(3)_{H}$&$SU(2)_{L}$ & $U(1)_Y$
\\
\hline
$Q_0$ & ${\mathbf 3}$& ${\mathbf 1}$ & $0$
\\
$Q_L$& ${\mathbf 3}$& ${\mathbf 2}$ & $-1/2$
\\
$\bar{Q}_0$ & $\bar{\mathbf 3}$& ${\mathbf 1}$ & $0$
\\
$\bar{Q}_L$& $\bar{\mathbf 3}$& $\bar{\mathbf 2}$ & $1/2$
\\
\hline
\end{tabular}
\end{center}
\label{tab:deformed}
\end{table}%

As a first step in this attempt, we consider a strongly coupled theory based on an $SU(3)$ supersymmetric QCD with three-flavors, $(Q_i, \bar{Q}_i)$
($i=1-3$), having a deformed moduli space below the dynamical scale
$\Lambda_H$\,\cite{Seiberg:1994bz}.
The charge assignments of these fields are given in Table\,\ref{tab:deformed}.
We allow tree-level interactions between the strongly interacting spectators and $H_u$,
$H_d$, and $X$ at high energies which are given by,
\begin{eqnarray}
 W_{\rm tree} =  \lambda_u H_u {Q}_L \bar Q_0
 +
  \lambda_d H_d \bar Q_L {Q}_0
  +
  \lambda_X X ( \bar{Q}_L Q_L+ \bar{Q}_0 Q_0 ) \ ,
\end{eqnarray}
where the summation of the gauge indices are understood.
We have taken a common coupling constant of $X$ to $Q_{L,0}$ for simplicity.
The $Q$'s become massless in the limit $\l_X\to 0$.
It should be noted that we have assumed that the $\mu$-term of the elementary Higgs doublets
is absent from the superpotential.%
\footnote{This can be enforced by appropriate symmetries such as a global $U(1)$ symmetry.}

Below the dynamical scale of $SU(3)_H$, $\L_H$, the light degrees of freedom are composite mesons and baryons which are related to the
elementary fields by,
\begin{eqnarray}
M^i_j &\simeq& \frac{1}{N_{\rm NDA}} \frac{(Q^i_a\bar{Q}_j^a)}{\L_H}\  ,\nonumber\\
B &\simeq& \frac{1}{N_{\rm NDA}}
\frac{ Q^{i_1}_{[a_1} Q^{i_2}_{a_2} Q^{i_3}_{a_3]}}{\L_H^2} \ , \nonumber\\
\bar{B} &\simeq&  \frac{1}{N_{\rm NDA}}
\frac{\bar{Q}_{i_1}^{[a_1} \bar{Q}_{i_2}^{a_2} \bar{Q}_{i_3}^{a_3]} }{\L_H^2}
\ .
\end{eqnarray}
The contraction of the gauge indices $a$ are understood.
Here, we have again used naive dimensional analysis and assumed that the above composite
fields have canonical kinetic terms.
In terms of these composite fields, the low energy effective superpotential is given by,
\begin{eqnarray}
\label{eq:Weff}
 W_{\rm eff}  &\simeq&  \frac{\l_u}{N_{\rm NDA}} \L_H H_u {\cal H}_d
+   \frac{\l_d}{N_{\rm NDA}}\L_H H_d {\cal H}_u +  \frac{\sqrt{3}\l_X}{N_{\rm NDA}}\L_H X M_0
\nonumber\\
&&+ \frac{N_{\rm NDA}}{\L_H}{\cal X}
\left(\det M+\frac{\L_H}{N_{\rm NDA}}B \bar{B} - \frac{\L_H^3}{N_{\rm NDA}^3}\right)\ ,
\end{eqnarray}
where ${\cal X}$ is a Lagrange multiplier field which enforces the deformed moduli constraint between the mesons and baryons.
We have neglected non-calculable $O(1)$ corrections to the coupling constants.
In the above expression, we have decomposed the meson fields
into two $SU(2)_L$ doublets $({\cal H}_u, {\cal H}_d)$,
one $SU(2)_L$ triplet ${\cal T}$, and two singlets $M_0 = \tr[M/\sqrt{3}]$
and $M_8 = \tr[\lambda_8 M]$ where $\l_8$ is the eighth Gell-Mann matrix of $SU(3)$.%
\footnote{Here, we take the normalization of the Gell-Mann matrix $\lambda_i$ $(i=1-8)$
to be $\tr[\l_i \l_j] = \delta_{ij}$.}

By expanding the meson and baryon fields around a solution of the deformed moduli constraint,
 \begin{eqnarray}
M_0 &\simeq&\sqrt{3} \frac{\L_H}{N_{\rm NDA}} + \delta M_0\ , \nonumber\\
{\cal X} & \simeq & - \lambda_X X\ ,
\end{eqnarray}
and all other fields are taken to be zero, the above superpotential is reduced to
\begin{eqnarray}
  W_{\rm eff}  &\simeq&  \frac{\l_u}{N_{\rm NDA}} \L_H H_u {\cal H}_d
+   \frac{\l_d}{N_{\rm NDA}}\L_H H_d {\cal H}_u
 + \lambda_X X{\cal H}_u{\cal H}_d\nonumber\\
&&+ \frac{\lambda_X}{2} X {\cal T}^2
+ \frac{\lambda_X}{2} X M_8^2
 - \lambda_X X \d M_0^2
- \lambda_X X B\bar{B}+ \cdots\ ,
\end{eqnarray}
where the the ellipses denote higher dimensional operators which are irrelevant
for our discussion.
As a result, we find that the composite mesons and baryons obtain masses
of about $M_H = \l_X M_X$, while the elementary Higgs doublets
have Dirac mass mixing terms together with their composite partners  ${\cal H}_{u,d}$.

Now, let us assume that $\l_{u,d}\L_H/N_{\rm NDA}\ll M_H$.
In this case, we may integrate out the composite mesons and baryons
at the scale of $M_H$ which leads to an effective $\mu$-term,
\begin{eqnarray}
\label{eq:effmu}
 W\simeq - \frac{\l_u\l_d}{N_{\rm NDA}^2}\frac{\L_H^2}{\l_X X} H_u H_d\ .
\end{eqnarray}
or more specifically the $\mu$-parameter is
\begin{eqnarray}
  \mu &\simeq& - \frac{\l_u\l_d}{N_{\rm NDA}^2} \frac{\L_H^2}{M_H}\ .
\end{eqnarray}
with $O(1)$ ambiguities.
It should also be noted that the effective $\mu$-term in Eq.\,(\ref{eq:effmu})
also leads to the  supersymmetry breaking Higgs mixing mass parameter,
\begin{eqnarray}
\label{eq:effB}
   B &\simeq& x\, M_X  = \frac{x}{\lambda_X} M_H\ .
\end{eqnarray}
Thus, we find that an appropriately sized $\mu$-term and $B$-term are generated by the effects of the strongly interacting spectator fields.

\begin{figure}[tb]
\begin{minipage}{.49\linewidth}
\begin{center}
\footnotesize${\hspace{1.2cm} [N_{NDA}=4\pi]}$
  \includegraphics[width=.9\linewidth]{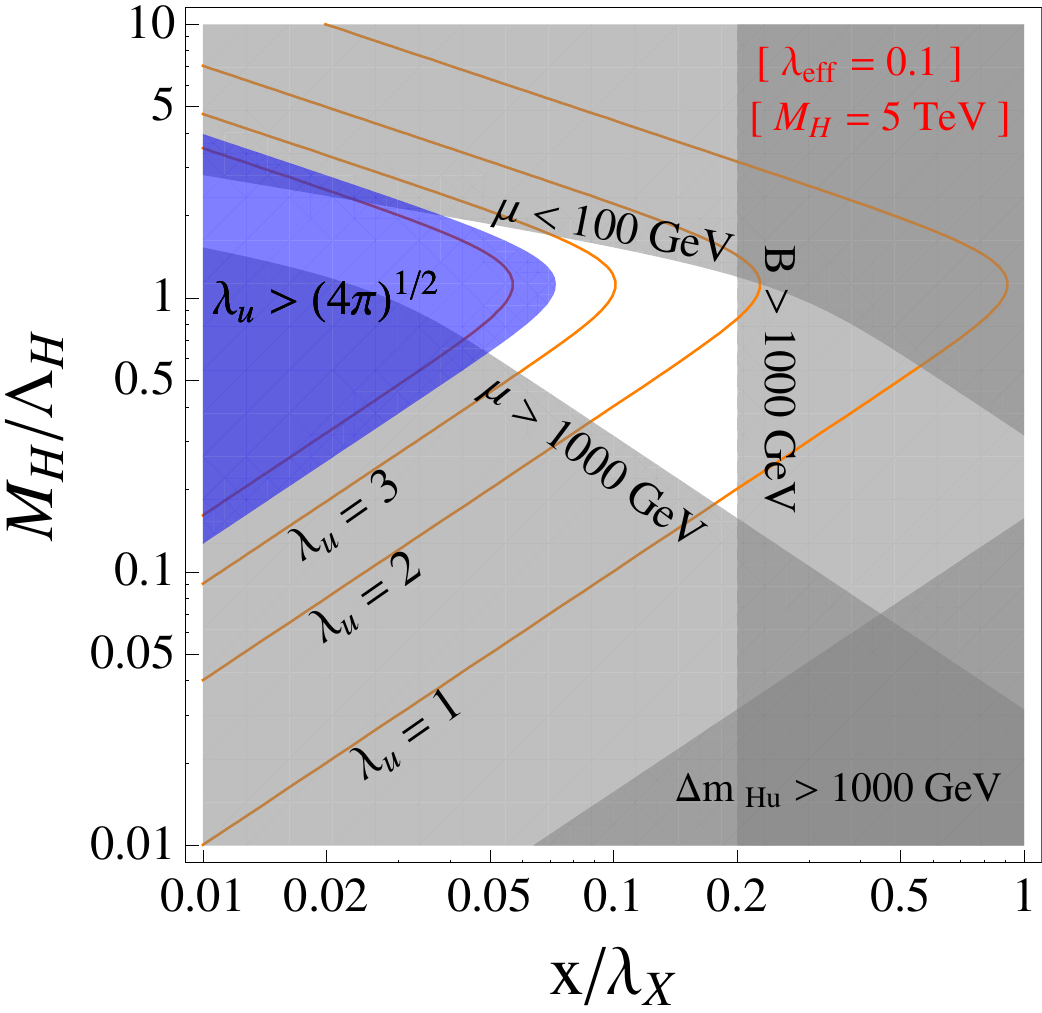}
 \end{center}
 \end{minipage}
 \begin{minipage}{.49\linewidth}
\begin{center}
\footnotesize${\hspace{1.2cm} [N_{NDA}=4\pi]}$
  \includegraphics[width=.9\linewidth]{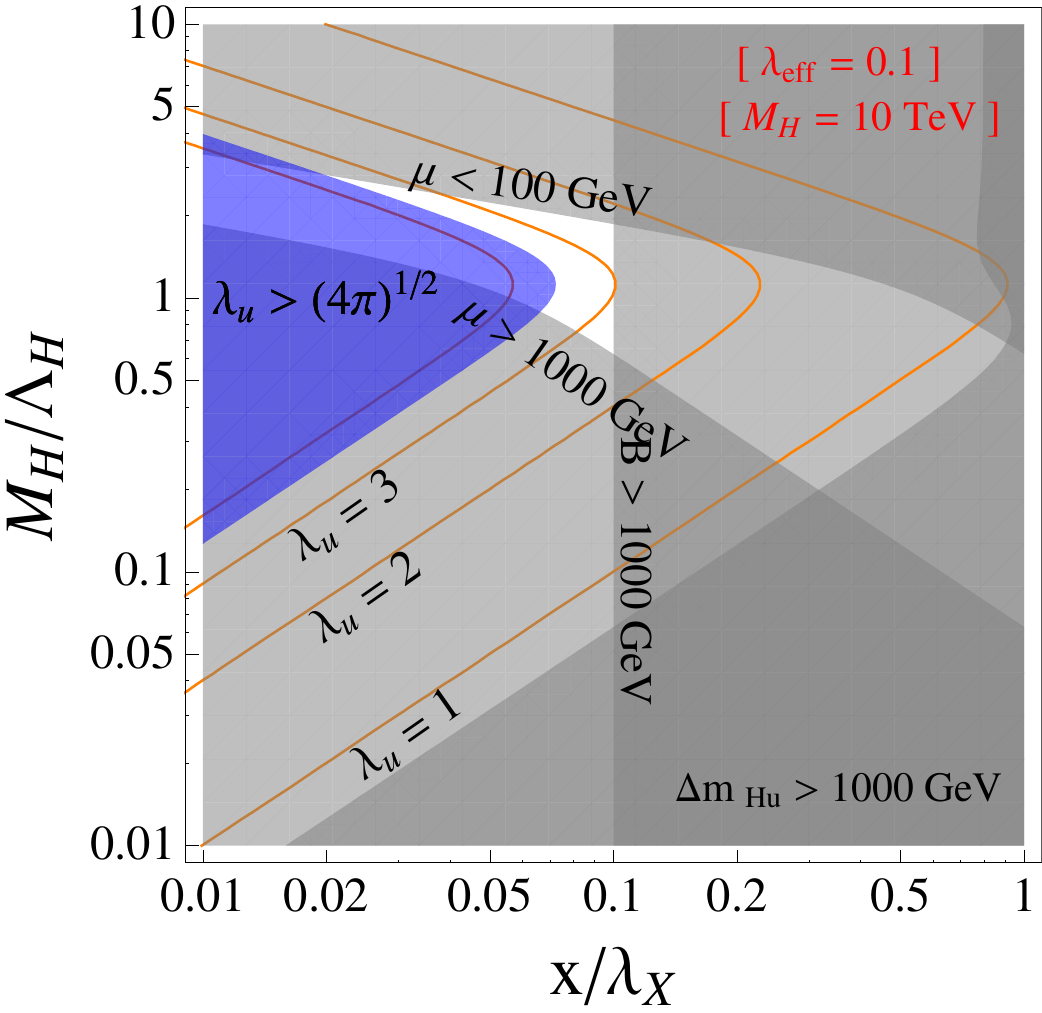}
 \end{center}
 \end{minipage}
\caption{\sl \small
Contour plots of $\lambda_u$ which realizes $\lambda_{\rm eff} = 0.1$
for $M_H = 5$\,TeV (left) and for $M_H=10$\,TeV (right) for $N_{NDA}=4\pi$.
The gray shaded regions are disfavored because $\mu$, $B$,  ${\mit \D}m_{H_u}$ are too large
or $\mu$ is too small. we have also assumed $\lambda_u = \lambda_d$.
}
\label{fig:lambdau2}
\end{figure}

Let us return to the effective quartic term of the Higgs doublets.
In addition to the effective quartic term in Eq.\,(\ref{eq:K4}), the quartic coupling of elementary Higgs doublets also receives additional contributions through its mixing with the composite Higgs,  ${\cal H}_{u,d}$.
That is, the composite Higgs doublets have an effective quartic term%
\footnote{
It should be noted that there is no effective quartic K\"ahler potential of $\cal H$'s
proportional to $(\l_X X)^{-2}$,
although the model includes massless fields in the limit of $\lambda_X \to 0$.
This is due to the fact that all the low energy interactions of those light composites states
are proportional to $\lambda_X X$ (see Eq.\,(\ref{eq:Weff})).
}
\begin{eqnarray}
K\simeq\frac{N_{\rm NDA}^2}{\L_H^2}{\cal H}_u^\dagger {\cal H}_u {\cal H}_u^\dagger {\cal H}_u \ .
\end{eqnarray}
The mixing of $H_u$ and ${\cal H}_u$ takes the above contribution to the K\"ahler potential and turns it into
\begin{eqnarray}
\label{eq:K42}
K\simeq\frac{\l_u^4\L_H^2} {N_{\rm NDA}^2|{\l_X}X|^4}{H}_u^\dagger {H}_u {H}_u^\dagger { H}_u \ .
\end{eqnarray}
As a result, we obtain
\begin{eqnarray}
\label{eq:lambdau2}
 \lambda_{\rm eff} \simeq \frac{16\l_u^4}{N_{\rm NDA}^2}\frac{\L_H^2}{M_H^2} \frac{x^2}{\l_X^2}\sin^4\beta\ ,
\end{eqnarray}
which is larger than that found in Eq.\,(\ref{eq:K4}) for $M_H \lesssim \L_H$.
The soft mass squared of $H_u$ also receives an additional contribution which is larger than the contribution found in Eq.\,(\ref{eq:mHu}). This contribution arise from the effective K\"ahler potential,
\begin{eqnarray}
\label{eq:K22}
K\simeq\frac{\l_u^2\L_H^2} {N_{\rm NDA}^2|{\l_X}X|^2}{H}_u^\dagger {H}_u \ ,
\end{eqnarray}
and leads to
\begin{eqnarray}
\label{eq:mHu2}
{ \mit \D} m_{H_u}^2 \simeq
  \frac{\l_u^2}{N_{\rm NDA}^2}\frac{x^2}{\l_X^2} \L_H^2\ .
\end{eqnarray}

In Fig.\,\ref{fig:lambdau2}, we show contours of $\lambda_u$ which give $\lambda_{\rm eff} = 0.1$, where we have assumed that $\l_{\rm eff}$ is given by a sum of the contributions in Eqs.\,(\ref{eq:K4})
and (\ref{eq:lambdau2}).
The shaded regions are disfavored because $\mu$, $B$ or ${\mit \D}m_{H_u}$ are larger than $1$\,TeV
or $\mu$ is smaller than $100$\,GeV, however, the boundaries of these regions are not exact since there are $O(1)$ ambiguities.
These figures show that an acceptable $\mu$-term and sufficiently large effective quartic coupling
constants are obtained if $M_H/\L_H \simeq 1$ and $\lambda_{u,d} \simeq 2-4$.

It should also be noted that the spectator fields charged under the strongly coupled gauge group of the hidden sector also generate $A$-terms and the wrong Higgs coupling $A$-terms\,\cite{Komargodski:2008ax}.
In our case, the $A$-terms are generated via the effective K\"ahler potential in Eq.\,(\ref{eq:K22})
which leads to
\begin{eqnarray}
 {\cal L} \simeq  \frac{\l_u^2 \L_H^2}{N_{\rm NDA}^2 M_H} \frac{x}{\l_X}
F_{H_u}^\dagger H_u + h.c.\ .
\end{eqnarray}
Therefore, the generated $A$-terms are suppressed by an additional factor of $(\l_u/N_{\rm NDA})^2$ as compared to the $B$-term in Eq.\,(\ref{eq:effB}).
The wrong Higgs couplings are generated though an effective
K\"ahler potential such as,%
\footnote{Here, we are assuming $\L_H\sim M_H$ which is favored in the above discussion
(see Fig.\,\ref{fig:lambdau2}).}
\begin{eqnarray}
 K \simeq  \frac{\l_u\l_d}{N_{\rm NDA}^2 }\frac{|\l_X X|^2}{\L_H^4}(\l_X X)^\dagger
 H_u (D^2 H_d)\ .
\end{eqnarray}
As a result, we obtain an effective operator leading to wrong Higgs coupling $A$-terms,
\begin{eqnarray}
 {\cal L} \simeq  \frac{\l_u\l_d \L_H^4}{N_{\rm NDA}^2 M_H^3} \frac{x^2}{\l_X^2}
F_{H_d} H_u + h.c.\ .
\end{eqnarray}
Thus, the wrong Higgs coupling $A$-terms is further suppressed by a factor of $x/\l_X$.
.

Next, we discuss the effects on the MSSM Yukawa coupling from the mixing of the elementary and composite Higgs bosons.
Since the elementary Higgs doublets mix with the composite Higgs doublets with a mixing angle
 $\varepsilon_{u,d} \simeq \l_{u,d}/N_{\rm NDA}\cdot \L_H/ M_H$,
 the normalizations of the Yukawa coupling constants above the threshold $M_H$
 are different from those in the MSSM.
For example, the top Yukawa coupling above the threshold $M_H$ is given by,
\begin{eqnarray}
y_{t}^{H} \simeq (1 + \varepsilon_u^2)^{1/2} y_t^{L}\ ,
\end{eqnarray}
where $y_t^{L}$ is determined by the top quark mass.
Clearly the high energy Yukawa coupling constant is larger than the low energy one. This could exasperate the Landau pole problem of the top Yukawa coupling.
For typical parameters found in Fig.\,\ref{fig:lambdau2}, however,
the effects of the Higgs mixing on the Yukawa couplings is quite small,
\begin{eqnarray}
\frac{ y_t^{H} -y_t^{L} }{y_{t}^L} \simeq \frac{1}{2} \varepsilon_u^2 \simeq 0.03\times
 \left( \frac{4\pi}{N_{\rm NDA}}\right)^2\left( \frac{\l_u}{3}\right)^2\frac{\L_H^2}{M_H^2}\ .
\end{eqnarray}
Thus, the mixings between the elementary and
composite Higgs doublets have a minor effect on the Landau problem.

Before closing this section, we comment on the stability of the baryons.  Although the low energy superpotential potential respects $U(1)_B$, this symmetry may be broken by Plank suppressed operators. As we will see below, these Plank suppressed operators are sufficient to guarantee that the baryons decay before BBN, thanks to the large anomalous dimensions of the strongly interacting fields.

\subsection{Constraints from Electroweak Precision Measurements}
\label{sec:ST}
As we have seen above, the effective quartic term and  $\mu$-term
are successfully generated from the spectator fields.
Interactions with the Higgs boson similar to those presented above
often contribute to the electroweak precision parameters which are
severely constrained by precision measurements.

Below the dynamical scale,
the most important higher dimensional operator which contributes to the electroweak precision
parameters\,\cite{Peskin:1990zt} is given by,
\begin{eqnarray}
K\simeq\frac{\l_u^4} {N_{\rm NDA}^2 \Lambda_H^2}{H}_u^\dagger {H}_u {H}_u^\dagger { H}_u \ ,
\end{eqnarray}
where we have assumed $\L_H \simeq M_H$.
This higher dimensional operator leads to the effective Lagrangian
\begin{eqnarray}
{\cal L}\simeq \frac{\l_u^4}{N_{\rm NDA}^2 \L_H^2} \sin^4\beta|h^\dagger D_\mu h|^2\ ,
\end{eqnarray}
which contributes to the $T$-parameter (see Ref.\,\cite{Carena:2009gx} for more extensive studies.).
As a result,
the
contribution to the $T$-parameter from the strongly interacting spectator fields is roughly estimated to be
\begin{eqnarray}
| T | \simeq \frac{v^2}{\alpha} \frac{\l_u^4}{N_{\rm NDA}^2 \L_H^2} \sin^4\b
\simeq 0.08 \times
\left(\frac{4\pi}{N_{\rm NDA}}\right)^2
\left(\frac{\l_u}{3}\right)^4
\left(\frac{5\,\rm TeV}{\L_H}\right)^2\sin^4\b\ ,
\end{eqnarray}
where we have again used naive dimensional analysis.
We have also used $v\simeq 174.1$\,GeV and $\alpha \simeq 1/129$ in the above expression.
Notice that we cannot determine the sign of $T$ when using naive dimensional analysis.
As we will see in what follows, the analysis in the perturbative limit shows that
the contribution from the spectator fields is positive.

In addition to the contributions to the $T$ parameters, there are operators which
contribute to the $S$-parameter such as,
\begin{eqnarray}
 K  \simeq\frac{\l_u^2} {N_{\rm NDA}^2 \Lambda_H^2} (\nabla^{\dagger 2} H_u^\dagger e^{-2V})
  (\nabla^2 H_u)\ ,
\end{eqnarray}
where $\nabla$'s denote the gauge covariant superspace derivatives.
This higher dimensional operator leads to an effective operator,
\begin{eqnarray}
{\cal L}\simeq \frac{\l_u^2gg'}{N_{\rm NDA}^2 \L_H^2} \sin^2\b( h^\dagger W_{\mu\nu} h) B^{\mu\nu} \ ,
\end{eqnarray}
where $W$, $B$ denote the gauge field strengths of $SU(2)_L$ and $U(1)_Y$
and $g$, $g'$ the corresponding gauge coupling constants, respectively.
As a result, the contribution to the $S$-parameter from the spectator fields is estimated to be
\begin{eqnarray}
|S| \simeq \frac{8 s_Wc_W v^2gg'}{\alpha}  \frac{\l_u^2}{N_{\rm NDA}^2 \L_H^2} \sin^2\b
\simeq 0.007 \times
\left(\frac{4\pi}{N_{\rm NDA}}\right)^2
\left(\frac{\l_u}{3}\right)^2
\left(\frac{5\,\rm TeV}{\L_H}\right)^2\sin^2\b\ ,
\end{eqnarray}
where $s_W$ and $c_W$ are the sine and cosine of the weak mixing angle.

Therefore, by comparing these contributions with the current constraint
(for $m_h = 120$\,GeV)\cite{Ludwig:2010vk};
\begin{eqnarray}
S = 0.02\pm 0.11\ ,\cr
T= 0.05\pm 0.12\  ,
\end{eqnarray}
we find that the strong dynamics of the spectator fields for $\L_H \simeq 5-10$\,TeV
give contributions that are well within the constraints of electroweak precision measurements.

\subsection{Perturbative Analysis}
Before closing this section, let us consider the effective quartic coupling
constant and the contributions to the electroweak precision parameters
in the limit of weak interactions.
In this case, the spectator sector is well described by the elementary $Q$'s,
and we can calculate the effective quartic coupling constant and the electroweak
precision parameters perturbatively.%
\footnote{
Similar perturbative analysis has was done in Ref. \,\cite{MO,Babu:2008ge,Martin:2009bg,Endo:2011mc,Evans:2011uq} where Higgs couples to additional matter with positive supersymmetry breaking masses.
}

At the one-loop level, the effective quartic coupling constant
is given by,
\begin{eqnarray}
 \l_{\rm eff}&\simeq& \frac{\l_u^4}{32\pi^2} \frac{\sin^4\b}{x_\l(1-x_\l^2)}
 (-2 x_\l (3-9 x_\l^2+8x_\l^4)
\cr
&& +  3 (-1 + 2 x_\l)(1-x_\l^2)^2 \log (1- x_\l)
  +  3 (1 + 2 x_\l)(1-x_\l^2)^2 \log (1+ x_\l))\cr
  &\simeq& \frac{\l_u^4}{16\pi^2} x_\l^2\,\sin^4\b \quad (x_\l \ll 1)\ ,
\end{eqnarray}
where we have defined $x_\l = x/\lambda_X$.
The advantage of this perturbative model is that
we can calculate the sign of the effective quartic term.
As we see from the results, the effective quartic coupling obtained
in the perturbative analysis is positive, and so enhances the Higgs boson mass.
This is an encouraging result for our model even though we are interested in the strongly
coupled regime where the perturbative calculation is no longer reliable.

Similarly, we can also calculate the contributions to the $S$ and $T$ parameters, from the strongly interacting spectator fields
perturbatively.
At the one-loop level, the scalars contribute
\begin{eqnarray}
S_s=\frac{N_c}{30\pi}\frac{\lambda_u^2v^2}{M_H^2}\frac{1-4x_\l^2-(1+x_\l^2)\sin\beta\cos\beta}{(1-x_\l^2)^2}\ ,
\end{eqnarray}
and
\begin{eqnarray}
&& T_s=\frac{N_c}{32\pi c_W^2s_W^2}\frac{\lambda_u^4v^4}{M_H^2M_Z^2}\left(\frac{1}{15}F_p(x_\l)(\sin\beta+\cos\beta)^4 +\frac{1}{3}F_m(x_\l)(\sin\beta-\cos\beta)^4\right.\\
\nonumber && \quad\quad\quad\quad\quad\quad\quad\quad\quad\quad \left. +F_{mp}(x_\l)(\sin^2\beta-\cos^2\beta)^2\right)\ ,
\end{eqnarray}
where
\begin{eqnarray}
&& F_p(x)=\frac{15x^4+14x^2+3}{(1-x^2)^3}\ , \\
&& F_m(x)=\frac{1}{1-x^2}\ ,\\
&& F_{mp}(x)=\frac{2x-4x^3+\frac{2}{3}x^5-(1-x^2)^3\ln\left(\frac{1+x}{1-x}\right)}{x^3(1-x^2)^2} \ .
\end{eqnarray}
The fermion contributions to the the $S$ and $T$ parameters are
\begin{eqnarray}
T_f=\frac{N_c}{480\pi c_W^2s_W^2}\frac{\lambda_u^4v^4}{M_H^2M_Z^2}(13+2\cos\beta\sin\beta-8\cos^2\beta\sin^2\beta )\ ,
\end{eqnarray}
and
\begin{eqnarray}
S_f=\frac{N_c}{30\pi}\frac{\lambda_u^2v^2}{M_H^2}(4-7\cos\beta\sin\beta)\ .
\end{eqnarray}
The total contributions are just
\begin{eqnarray}
T=T_s+T_f \ , \\
S=S_s+S_f\ ,
\end{eqnarray}
which gives a result similar to
the estimations made based on naive dimensional analysis in the previous
section with $\l_u  = O(1)$.%
\footnote{For a more detailed analysis see Appendix \ref{STU}.}.

\section{Strong Conformal Dynamics of the Spectators}
\label{sec:conformal}
\begin{table}[t]
\caption{\sl \small
The strongly coupled sector is charged under $SU(3)_H$ which is based on supersymmetric
QCD with six-flavors.
We embed the Standard Model gauge groups $SU(3)_c\times SU(2)_L\times U(1)_Y$ into
the subgroup of the maximal global symmetry $U(6)\times U(6)$.
}
\begin{center}
\begin{tabular}{|c|c|c|c|c|c|}
\hline
&$SU(3)_{H}$&$SU(3)_c$&$SU(2)_{L}$ & $U(1)_Y$
\\
\hline
$Q_0$ & ${\mathbf 3}$& ${\mathbf 1}$& ${\mathbf 1}$ & $0$
\\
$Q_L$& ${\mathbf 3}$& ${\mathbf 1}$& ${\mathbf 2}$ & $-1/2$
\\
$Q_{\bar D}$& ${\mathbf 3}$& $\bar {\mathbf 3}$& ${\mathbf 1}$ & $1/3$
\\
$\bar{Q}_0$ & $\bar{\mathbf 3}$& ${\mathbf 1}$& ${\mathbf 1}$ & $0$
\\
$\bar{Q}_L$& $\bar{\mathbf 3}$& ${\mathbf 1}$& $\bar{\mathbf 2}$ & $1/2$
\\
$\bar{Q}_{\bar D}$& $\bar {\mathbf 3}$& ${{\mathbf 3}}$& ${\mathbf 1}$ & $-1/3$
\\
\hline
\end{tabular}
\end{center}
\label{tab:conformal}
\end{table}
In the previous section, we have constructed a model with strongly interacting spectators based on supersymmetric QCD with a quantum deformed moduli space.
There, we showed that the $\mu$-term and a
large effective quartic term could be generated as a result of the spectator fields.
In the model of the previous section, however, there are several unsatisfactory features;
\begin{itemize}
\item The needed coupling constants $\l_{u,d}$ are rather large (see Fig.\,\ref{fig:lambdau2}).
\item The ratio between $M_H$ and $\L_H$ needs to be close to one
 (see Fig.\,\ref{fig:lambdau2}).
\item The matter content of the strongly interacting sector is not consistent with Grand Unification
(see Table\,\ref{tab:deformed}).
\end{itemize}
In this section, we show that these unsatisfactory features can be
solved simultaneously by simply extending this model into the conformal window.

\subsection{Model Near the Conformal Fixed Point}
In the previous section, we have considered a model with strongly interacting spectators
based on an $SU(3)$ supersymmetric QCD with three-flavors.
Here, we make a simple extension of this theory to six-flavors
with the charge assignments given in Table\,\ref{tab:conformal}.
As we see from the table, the matter content of this sector
is now consistent with an $SU(5)$ GUT.
With this simple extension, we now assume a tree-level superpotential of
\begin{eqnarray}
 W_{\rm tree} =  \lambda_u H_u {Q}_L \bar Q_0
 +
  \lambda_d H_d \bar Q_L {Q}_0
  +
  \lambda_X X ( \bar{Q}_{\bar D} Q_{\bar D}+\bar{Q}_L Q_L+ \bar{Q}_0 Q_0) \ .
\end{eqnarray}
For this theory the $\mu$ and $B$-terms as well as the effective quartic term of the Higgs boson
are generated in the same way they were in the previous sections.%
\footnote{
Strictly speaking, in order for this model to flow to the deformed moduli model
discussed in the previous section, we need to assume $M_{Q_D}> M_{Q_{L,0}}$.
Although this is an assumption, it should arise naturally as a result
of the RG flows of GUT coupling constant $\lambda_X$,
because $Q_D$ is charged under $SU(3)_c$.}

\begin{figure}[tb]
\begin{center}
  \includegraphics[width=.5\linewidth]{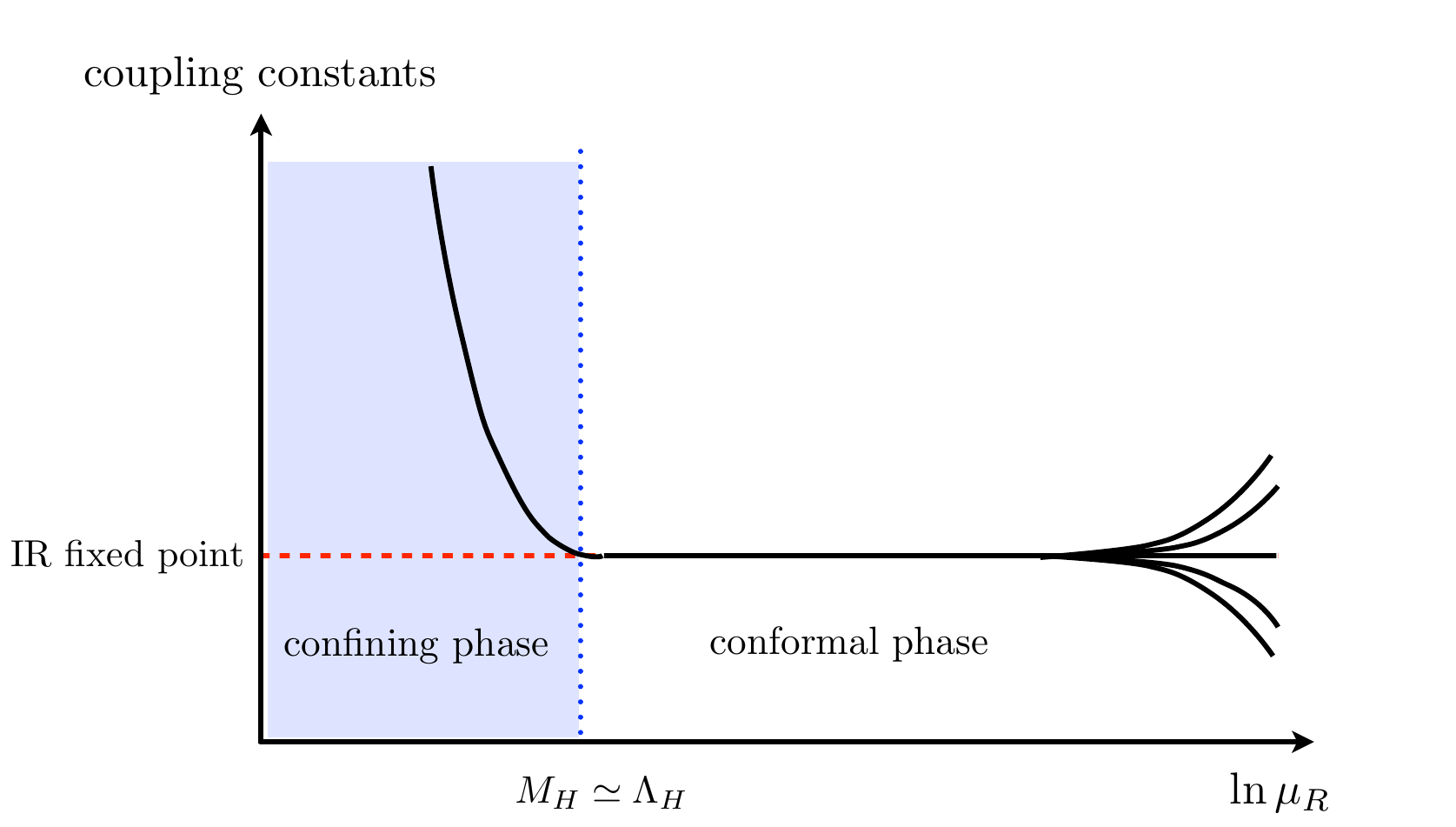}
\caption{\sl \small
An illustrative picture of the RG flow of the conformal sector.
The parameter $\mu_R$ denotes the renormalization scale.
The conformal symmetry is broken explicit by the mass term $M_H = \l_X M_X$.
Below this scale, the strongly interacting sector flows into a confining phase as discussed in the previous section.
}
\label{fig:flow}
\end{center}
\end{figure}

It should be noted, however, that the $SU(3)_H$  supersymmetric gauge theory
with six-flavors is in the conformal window\,\cite{Seiberg:1994pq}, and hence,
the coupling constants flow to an infrared fixed point
in the limit $\l_X \to 0$ (i.e. $M_H \to 0$).
Thus, if the coupling constants at the high energy scale are
in the vicinity of the infrared fixed point, the coupling constants will subsequently
flow to the fixed point.
The conformal symmetry is eventually broken by the explicit mass term $M_H = \l_X M_X$
below which the strongly coupled sector flows to a confining phase and behaves as discussed in the previous section.
In this way, the effective quartic term as well as the $\mu$ and $B$ terms
are generated in this model with strongly coupled conformal dynamics.
In Fig.\,\ref{fig:flow}, we show an illustrative picture of
the RG flow of the conformal sector.

Now, let us estimate the coupling constants at the infrared fixed point.
Initially, we will neglect the MSSM couplings.
(We discuss the renormalization group flow including the MSSM coupling
constants in the next subsection.)
In this limit,  the one-loop anomalous dimensions of the matter fields
are given by,
\begin{eqnarray}
\gamma_{0} &=& \frac{1}{2\pi} \alpha_u - \frac{2}{3\pi} \alpha_{3'} \ ,  \,
\gamma_{L} = \frac{1}{4\pi} \alpha_d - \frac{2}{3\pi} \alpha_{3'} \ , \,
\gamma_{D} = - \frac{2}{3\pi} \alpha_{3'} \ , \,
\gamma_{H_u}= \frac{3}{4\pi} \alpha_u \ , \nonumber \\
\gamma_{\bar{0}} &=& \frac{1}{2\pi} \alpha_d - \frac{2}{3\pi} \alpha_{3'} \ , \,
\gamma_{\bar{L}} = \frac{1}{4\pi} \alpha_u - \frac{2}{3\pi} \alpha_{3'} \ , \,
\gamma_{\bar{D}} = - \frac{2}{3\pi} \alpha_{3'} \ , \,
\gamma_{H_d} = \frac{3}{4\pi} \alpha_d \ ,
\end{eqnarray}
where the subscripts of the anomalous dimensions corresponds to the ones
appearing in table\,\ref{tab:conformal} and
the $\alpha$'s are $\alpha_{3'} = g_{3'}^2/4\pi$ with $g_{3'}$ being
the gauge coupling constant of $SU(3)_H$,  $\alpha_u =\l_u^2/4\pi$
and $\alpha_u =\l_d^2/4\pi$.
In terms of the anomalous dimensions, the NSVZ beta function is given by,
\begin{eqnarray}
\frac{d}{d\ln\mu_R}\frac{1}{\a_{3'}} = \frac{1}{2\pi} \frac{9 -\sum(1- 2\gamma_{i})/2
-\sum(1- 2\gamma_{\bar{i}})/2
 }{1 - 3 \a_{3'}/2\pi }\ ,
\end{eqnarray}
where the summation is taken over all 6 flavors.
The beta functions of the Yukawa interactions are given by,
\begin{eqnarray}
\frac{d}{d\ln\mu_R}\a_u = 2\a_u(\gamma_{H_u} + \gamma_{\bar{u}}+ \gamma_{ 0} )\ ,
\quad
\frac{d}{d\ln\mu_R}\a_d =2\a_d( \gamma_{H_d} + \gamma_{u}+ \gamma_{\bar{0}} )\ .
\end{eqnarray}
By requiring that all the beta functions are vanishing,
we find three different infrared fixed points (the so-called Banks-Zaks approximation\,\cite{BZ});
\begin{eqnarray}
\label{eq:fixedpoint}
\begin{array}{llll }
(I)  : &   \displaystyle{{\l_u^2} = \frac{12\pi^2}{7}}\ , &
\displaystyle{{\l_d^2} = \frac{12\pi^2}{7}}\ ,&
\displaystyle{g_{3'}^2 = \frac{27\pi^2}{14}}\ ,  \\
(I\hspace{-.2em}I)  : &   \displaystyle{{\l_{u,d}^2} = \frac{3\pi^2}{2}}\ , &
\displaystyle{{\l_{d,u}^2} =0}\ , &
\displaystyle{g_{3'}^2 = \frac{27\pi^2}{16}}\ ,  \\
(I\hspace{-.25em}I\hspace{-.25em}I)  : &   \displaystyle{{\l_u^2} = 0}\ , &
\displaystyle{{\l_d^2} = 0}\ ,&
\displaystyle{g_{3'}^2 = \frac{3\pi^2}{2}}\ .  \\
\end{array}
\end{eqnarray}
Notice that the fixed points on the second line are only stable when either $\l_{u}$  or $\l_{d}$
are zero and the third fixed point is only stable for $\l_u = \l_d = 0$.

Interestingly, $\l_{u,d}$ and $g_{3'}$ take
rather large values for the stable fixed point ($I$),
\begin{eqnarray}
\label{eq:fixed}
 \l_u \simeq 4.1\ ,\quad   \l_d \simeq 4.1\ , \quad
  g_{3'} \simeq 4.4\ .
\end{eqnarray}
These large values are advantageous for generating a large effective quartic coupling for the Higgs boson as well as for generating natural values of the $\mu$ and $B$ term (see Fig.\,\ref{fig:lambdau2}).
Thus, if the strongly coupled sector approaches the fixed point
above the conformal breaking scale $M_H$,
the strongly coupled sector naturally leads to the desired coupling constants $\l_{u,d}$.

Another bonus of having a rather strongly interacting fixed point is that
the model predicts that $M_H$ should be close to the dynamical scale $\L_H$.
That is, the spectators become confined immediately after
the conformal symmetry breaking, since the gauge coupling constant
is already large at the fixed point.
Therefore, the model also importantly predicts
\begin{eqnarray}
 \L_H  \simeq M_H\ ,
\end{eqnarray}
which is needed as is discussed in the previous section.
As a result, we find that the conformal dynamics of the spectators provide
us a very attractive framework for producing parameters with the appropriate size.

Before closing this section, let us comment on the validity of the one-loop Banks-Zaks approximation.
As we have seen, the coupling constants at the fixed point
are rather large (see Eq.\,(\ref{eq:fixed})), and hence, the one-loop approximation
of the anomalous dimension seems less reliable.
To justify our use of the one-loop anomalous dimensions,
let us compare our one-loop approximation with the anomalous dimensions determined non-perturbatively
using $a$-maximization\,\cite{Intriligator:2003jj}.
As shown in appendix\,\ref{sec:a-max}, the anomalous dimensions determined by $a$-maximization are given by,
\begin{eqnarray}
 \gamma_{0} \simeq -0.18\ , \quad
 \gamma_{L} \simeq -0.23\ , \quad
  \gamma_{\bar D} \simeq -0.29\ , \quad
  \gamma_{H_u} \simeq 0.41\ ,
\end{eqnarray}
with $\gamma_{i} = \gamma_{\bar{i}}$ and $\gamma_{H_u} = \gamma_{H_d}$.
The anomalous dimensions estimated by the Banks-Zaks approximation
are, on the other hand,
 \begin{eqnarray}
 \gamma_{0} \simeq -0.11\ , \quad
 \gamma_{L} \simeq -0.21\ , \quad
  \gamma_{\bar D} \simeq -0.32\ , \quad
  \gamma_{H_u} \simeq 0.32\ .
\end{eqnarray}
The discrepancies between the one-loop approximation
and the non-perturbative determination are at most 30\%.
Therefore,  the  Banks-Zaks approximation provides us
moderately reliable results.

\subsection{Numerical Renormalization Group Flow}
\label{sec:RGE}
In the above discussion, we have neglected contributions from the MSSM coupling constants.
As we have seen, however, the coupling constants at the fixed point are quite large,
and hence, will affect the running of the MSSM coupling constants.
In particular, the beta function of the top Yukawa coupling
receives large positive contributions from the conformal Yukawa couplings
which drives the top Yukawa coupling constant large (small) at the
high (low) energy.
In particular, if the coupling constants of the conformal sector
reach the fixed point at some high energy scale, the top Yukawa coupling constant
is drastically suppressed in the low energy due to the renormalization group running.
The back reactions from the top Yukawa coupling onto the conformal coupling constants is
also not negligible.

\begin{figure}[tb]
\begin{center}
\begin{minipage}{.49\linewidth}
\begin{center}
  \includegraphics[width=.9\linewidth]{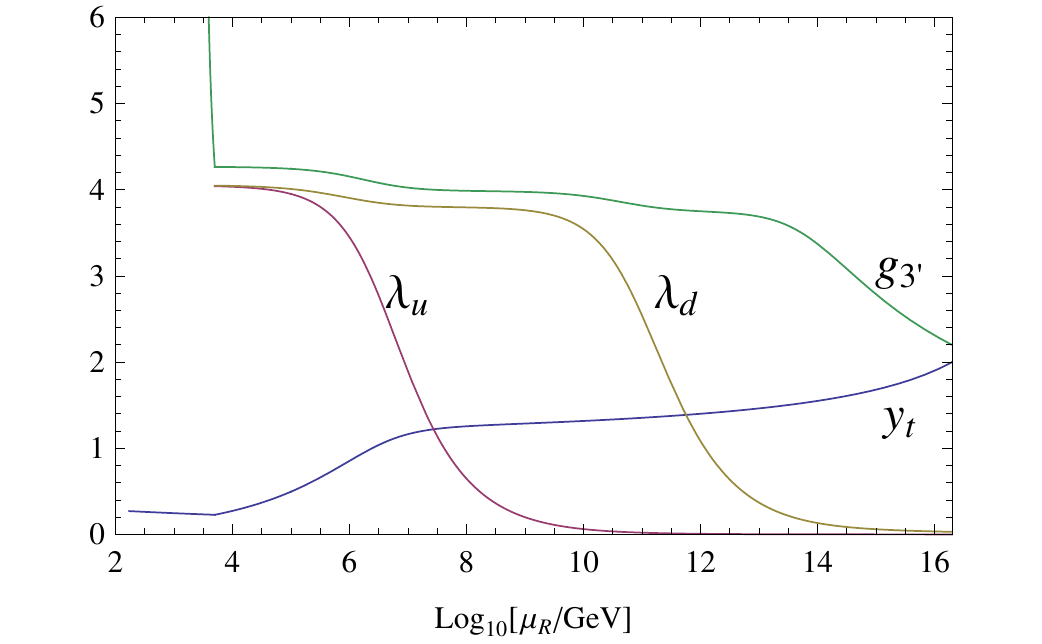}
\end{center}
 \end{minipage}
 \begin{minipage}{.49\linewidth}
 \begin{center}
  \includegraphics[width=.9\linewidth]{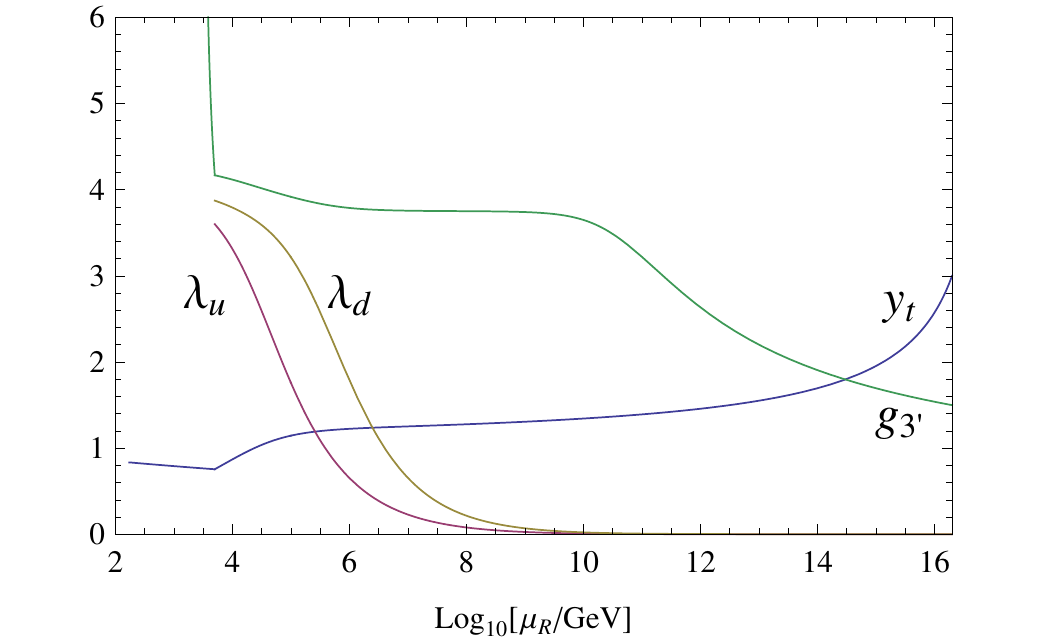}
 \end{center}
 \end{minipage}
\caption{\sl \small
Renormalization group running of the  coupling constants in the conformal sector
and the top Yukawa coupling constant.
Left) The running for $\lambda_u=3\times 10^{-4}$,
$\lambda_d=3\times 10^{-2}$,
$g_{3'}=2.2$, and $y_t = 2$ at the GUT scale
and $M_H= 5$\,TeV.
Right) The running for $\lambda_u=\lambda_d=1\times 10^{-3}$,
$g_{3'}=1.5$, and $y_t =3$ at the GUT scale
and $M_H= 5$\,TeV.
}
\label{fig:running}
\end{center}
\end{figure}

In Fig.\,\ref{fig:running}, we show typical renormalization group flows
of the conformal sector coupling constants and the top Yukawa coupling constant.	
The anomalous dimensions and the beta functions of the coupling
constants are given in appendix\,\ref{sec:RGEapp}.
The GUT scale values for the MSSM gauge coupling constants
are taken to be\footnote{ Because the anomalous dimensions of the $Q$'s are large, there two-loop order effect on the SM gauge couplings can be non-trivial. However, because of the quasi-fixed point nature of the running, this effect will be smaller than the order one ambiguities we have already neglected. } $g_{1,2,3}({M_{GUT}})\simeq1.1$.

The left panel of the figure exhibits all of the key features of the conformal sector.
Since $\l_{u,d}\ll 1$ at the GUT scale, the gauge coupling initially
runs to the fixed point at $g_{3'}^2 = 3\pi^2/2$ (i.e. the fixed point $(I\hspace{-.25em}I\hspace{-.25em}I)$
in Eq.\,(\ref{eq:fixedpoint})).
It remains at this fixed point values until $\l_d$ becomes sizable.
The conformal sector then proceeds to fixed point $(I\hspace{-.2em}I)$.
Once $\lambda_{u}$ becomes large, the theory moves onto the stable fixed point $(I)$.
Because $\l_u$ is rather large at fixed point $(I)$,
the top Yukawa coupling is driven to zero in the low energy.
At the scale $M_H$, the conformal sector is integrated out,
and running of top Yukawa coupling becomes like it is in the MSSM.

An interesting property of this renormalization group running is the altered
quasi-fixed point of the top Yukawa coupling\,\cite{Hill:1980sq}.
In the MSSM, the top Yukawa coupling has an infrared quasi-fixed point
value which predicts a top quark mass which is too heavy\,\cite{Brahmachari:1997yp}.
With the presence of the conformal sector, on the other hand,
the quasi-fixed point value
of the top Yukawa coupling is shifted to a much lower value.
Interestingly, we find that the quasi-fixed point of the top Yukawa coupling
is strongly correlated with the coupling constant $\l_u$ at $M_H$,
if $g_{3'}$ has already run to the fixed point  $(I\hspace{-.25em}I\hspace{-.25em}I)$
at some higher energy scale as in Fig.\,\ref{fig:running}.

\begin{figure}[tb]
\begin{center}
  \includegraphics[width=.5\linewidth]{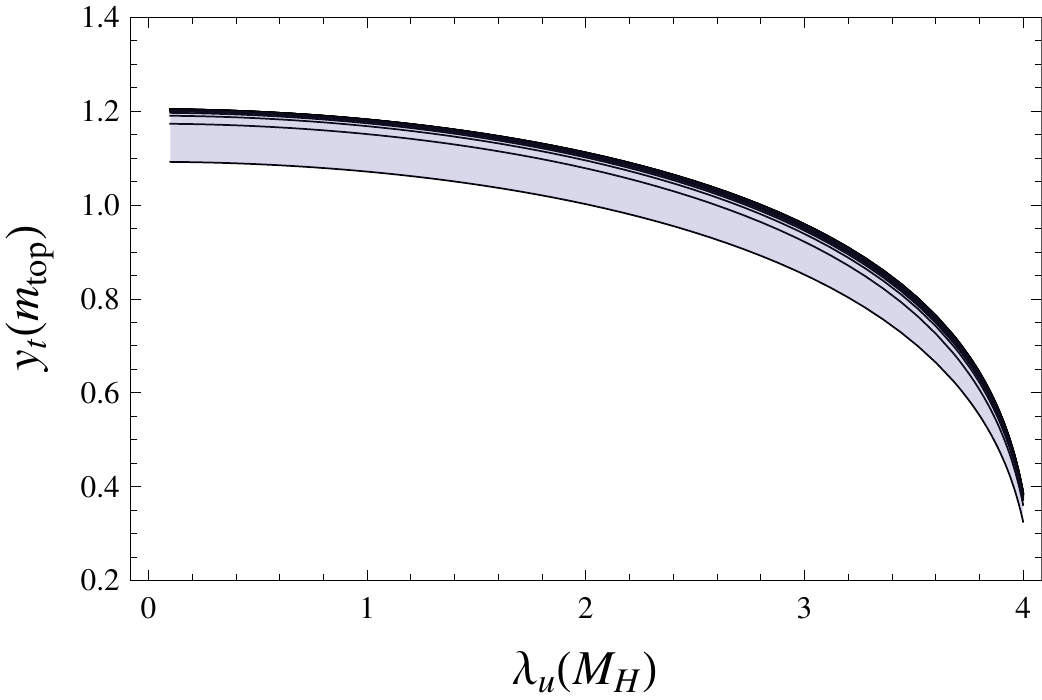}
\caption{\sl \small
The correlation between $\lambda_u(M_H)$ and $y_{t}(m_{\rm top})$.
Each line corresponds to GUT scale values of  $y_t = 0.5, 1,1.5\cdots 10$
from bottom to up respectively, but the lines for $y_t(M_{\rm GUT}) \geq 2$ are almost degenerate
with each other and cannot be resolved.
A value of $\lambda_u (M_H)\simeq 4$ corresponds
to the infrared fixed point $(I)$.
}
\label{fig:quasiFP}
\end{center}
\end{figure}

In Fig.\,\ref{fig:quasiFP}, we show the correlation between $\lambda_u(M_H)$
and $y_t(m_{\rm top})$ for various boundary condition at the GUT scale.
In the figure, we have taken $y_{t} = 0.5-10$ and $\l_u = 10^{-7}-10$ at the GUT scale.
We have also taken $\lambda_u = \lambda_d$  and $g_{3'}=1.5$ at the GUT scale
and $M_H = 5$\,TeV,
however, the quasi-fixed point is less sensitive to these choices as long as $g_{3'}$ reaches the fixed point  $(I\hspace{-.25em}I\hspace{-.25em}I)$ at some energy
scale much higher than $M_H$.

Fig. (\ref{fig:quasiFP}) shows that a wide range of GUT scale top Yukawa coupling constant (i.e. $y_t(M_{\rm GUT})=0.5-10$) are focused into a quasi-fixed point
for a given value of $\lambda_u(M_H)$.
As a result, the top Yukawa coupling constant determined from the
observed top quark mass $m_{\rm top} =173.2\pm 0.2$\,GeV\,\cite{Lancaster:2011wr}
\begin{eqnarray}
 y_{t}(m_{\rm top}) \simeq 0.94\times \left(\frac{m_{\rm top}}{173.2}\right)\ ,
\end{eqnarray}
predicts
\begin{eqnarray}
 \lambda_u(M_H)\simeq 3 \ .
\end{eqnarray}
This value is quite favorable for realizing a sufficiently large quartic coupling
as well as natural values for the $\mu$ and $B$-term (see Fig.\,\ref{fig:lambdau2}).

Finally, we comment of the stability of the baryons in this model. Because the anomalous dimensions of the hidden sector fields are quite large and relatively unchanging, the Plank suppressed operators like
\begin{eqnarray}
W=\frac{1}{M_P} 5_Q5_Q5_Q 10
\end{eqnarray}
with $5_Q=(\bar Q_{\bar D}, \bar Q_L)$, are enhanced as the theory is run down from the Plank scale.  This enhancement sufficiently destabilizes the baryons and this model is safe from BBN constraints.

\section{Conclusions and Discussions}
In this paper, we have proposed a new mechanism for increasing the mass of the lightest Higgs boson. The Higgs bosons mass is increased by coupling the Higgs boson to a strongly interacting conformal sector. The Higgs doublets are neutral under this additional gauge group, but feel its effects via
(semi-)perturbative Yukawa couplings.
As we have shown, the lightest Higgs boson mass suggested
by the ATLAS and CMS experiments can easily be
realized from this mechanism.
We have also constructed a model where
the $\mu$-term is successfully generated from the same dynamics.
Furthermore, we proposed a model in the conformal window
with appropriate values of the couplings at the fixed point for generating a quasi-natural Higgs boson mass of $125$ GeV.
This model portrays an interesting correlation
between the top Yukawa coupling constant and the Yukawa couplings of the conformal sector.

Finally, let us comment on other phenomenological studies
which we have left for future work.
First of all, enhancing the Higgs boson mass by coupling the MSSM Higgs to the conformal sector, as dicusssed in the text, works
even for large $\tan\b$.
Therefore, it is quite tempting to investigate whether
the observed deviation of the muon $g-2$ of about
$3.3\,\s$\,\cite{Hagiwara:2011af}
can be explained by coupling the MSSM with the conformal sector while still having a lightest Higgs
boson mass around $125$\,GeV.%
\footnote{
For recent model building which achieves
$g-2$ within $1\,\sigma$ and has a
$125$\,GeV Higgs boson mass, see Refs.\,\cite{Endo:2011xq,Endo:2011gy,Evans:2012hg}.
}

Another interesting phenomenological feature of our model is
the stop masses.
As we have discussed, the top Yukawa coupling
will be larger than in the MSSM for $\mu > \Lambda_H$.
Thus, the stop soft squared masses tend to receive larger negative contributions
from the renormalization group running, altering the typical MSSM stop spectrum. This suppression of the stop mass could alleviate some of the fine tuning of the MSSM.
The soft squared masses of the Higgs doublets are also affected
by the super conformal feature of the spectator fields it is coupled to.%
\footnote{The conformal nature of the spectator sector
could lead to a suppression of $m_{H_u}^2$\cite{Nelson:2000sn,Kobayashi:2001kz,Luty:2001jh,Ibe:2005pj,Schmaltz:2006qs,Cohen:2006qc} which could also mitigate some of the fine-tuning of the MSSM. }
These discussions are also left for future work.

\section*{Note added}
While completing this paper, an interesting article by
J.~J.~Heckman, P.~Kumar and B.~Wecht\,\cite{Heckman:2012nt}
was posted on arxiv
which also discussed the effects of coupling a strongly interacting sector to the Higgs, however, their model and effects of the strongly coupled sector are different from ours.

\section*{Acknowledgements}
This work is supported by Grant-in-Aid for Scientific research from the Ministry of Education, Science, Sports, and Culture (MEXT), Japan, No.\ 22244021 (T.T.Y.),
No.\ 24740151 (M.I),
 and also by World Premier International Research Center Initiative (WPI Initiative), MEXT, Japan.

\appendix
\section{$S$ and $T$ Parameters\label{STU}}
Here we give the full one-loop expressions for the $S$ and $T$  parameters.  The superpotential we consider here is
\begin{eqnarray}
W=Z( Q_{L} \bar  Q_L +\bar Q_0 Q_0) +\lambda_u H_u Q_L \bar Q_0 +\lambda_d H_d \bar Q_L Q_0\ .
\end{eqnarray}
where $Z=M_H+F_H\theta^2$. We further define the quark fields as
\begin{eqnarray}
Q_L=\left(\begin{array}{c} Q_\nu \\ Q_e \end{array}\right) \ , \quad & \quad \bar Q_L=\left(\begin{array}{c} \bar Q_\nu \\ \bar Q_e \end{array}\right)\ .
\end{eqnarray}
The mass matrix for the fermions is
\begin{eqnarray}
M_{\tilde \nu}=\left(\begin{array}{cc} M_H & \lambda_d v_d \\ \lambda_u v_u & M_H \end{array} \right)\ .
\end{eqnarray}
This matrix is diagonalized by
\begin{eqnarray}
\left(\begin{array}{c} Q_{\nu_2} \\ Q_{\nu_1}\end{array} \right)=\left(\begin{array}{cc} \sin\theta & -\cos\theta \\ \cos\theta & \sin\theta \end{array}\right) \left(\begin{array}{c} Q_{\nu} \\ Q_{0}\end{array}\right)\ , &
\left(\begin{array}{c} \bar Q_{\nu_2} \\  \bar Q_{\nu_1}\end{array} \right)=\left(\begin{array}{cc} \cos\theta & -\sin\theta \\ \sin\theta & \cos\theta \end{array}\right) \left(\begin{array}{c} \bar Q_{\nu} \\ \bar Q_{0}\end{array}\right)\ ,
\end{eqnarray}
where
\begin{eqnarray}
m_{\nu_1,\bar \nu_1}^2= \frac{1}{2} \left( 2M_H^2+v_u^2+v_d^2+\sqrt{(v_u^2-v_d^2)^2+4M_H^2(v_u+v_d)^2}\right)\ ,\cr
m_{\nu_2,\bar \nu_2}^2= \frac{1}{2} \left( 2M_H^2+v_u^2+v_d^2-\sqrt{(v_u^2-v_d^2)^2+4M_H^2(v_u+v_d)^2}\right)\ ,
\end{eqnarray}
and

\begin{eqnarray}
\cos\theta=\left(\frac{m_{\nu_1}^2-(M_H^2+v_d^2)}{m_{\nu_1}^2-m_{\nu_2}^2}\right)^{1/2}\ ,
\quad \sin\theta=\left(\frac{(M_H^2+v_d^2)-m_{\nu_2}^2}{m_{\nu_1}^2-m_{\nu_2}^2}\right)^{1/2}\ .
\end{eqnarray}

The mass matrix for the charges sleptons and its diagonalization matrix are
\begin{eqnarray}
M_L^2=\left( \begin{array}{cc} M_H & F_H \\
F_H & M_H \end{array} \right) \ ,\quad &\quad
R_L=\frac{1}{\sqrt{2}}\left( \begin{array}{rr} 1 & -1 \\
1 & 1 \end{array} \right)\ .
\end{eqnarray}
with mass eigenstates
\begin{eqnarray}
m_{e_1}^2=M_H^2+F_H\ ,\\
m_{e_2}^2=M_H^2-F_H\ .
\end{eqnarray}

The mass matrix for the neutral sleptons is
\begin{eqnarray}
M_0^2=\left( \begin{array}{cccc} |\lambda_u v_u|^2+M_H^2 & M_H(\lambda_u v_u +\lambda_d v_d) & 0 & F_H \\
M_H(\lambda_u v_u +\lambda_d v_d) & |\lambda_d v_d|^2+M_H^2& F_H & 0 \\
0 & F_H & |\lambda_u v_u|^2+M_H^2 & M_H(\lambda_u v_u +\lambda_d v_d) \\
F_H & 0 & M_H(\lambda_u v_u +\lambda_d v_d) & |\lambda_d v_d|^2+M_H^2
\end{array}\right)\ ,
\end{eqnarray}
in the basis $[Q_0, Q_\nu, \bar Q_\nu^\dagger , \bar Q_0^\dagger ]$.   The diagonalization matrix for this matrix is
\begin{eqnarray}
R_0=\frac{1}{\sqrt{2}}\left( \begin{array}{cccc} \cos\theta & -\sin\theta & \cos\bar \theta & -\sin\bar\theta \\
\sin\theta & \cos\theta & -\sin\bar\theta & -\cos\bar\theta \\
\cos\theta & -\sin\theta & -\cos\bar\theta & \sin\bar\theta \\
\sin\theta & \cos\theta & \sin\bar\theta & \cos\bar\theta
\end{array}\right)\ ,
\end{eqnarray}
where
\begin{eqnarray}
\sin\theta= \frac{1}{2}\left(\frac{M_1^2-M_2^2+(|\lambda_u v_u|^2-|\lambda_d v_d|^2)}{M_1^2-M_2^2}\right)^{1/2} \ , \cr
\cos\theta= \frac{1}{2}\left(\frac{M_1^2-M_2^2-(|\lambda_u v_u|^2-|\lambda_d v_d|^2)}{M_1^2-M_2^2}\right)^{1/2}\ , \cr
\sin\bar\theta= \frac{1}{2}\left(\frac{M_3^2-M_4^2+(|\lambda_u v_u|^2-|\lambda_d v_d|^2)}{M_3^2-M_4^2}\right)^{1/2} \ , \cr
\cos\bar\theta= \frac{1}{2}\left(\frac{M_3^2-M_4^2-(|\lambda_u v_u|^2-|\lambda_d v_d|^2)}{M_3^2-M_4^2}\right)^{1/2}\ .
\end{eqnarray}
and the mass eigenstates are
\begin{eqnarray}
\!\!\!\!\! m_{0_1}^2=M_H^2+\frac{1}{2}(|\lambda_u v_u|+|\lambda_u v_u|^2)^2+\frac{1}{2}\sqrt{(|\lambda_u v_u|^2-|\lambda_u v_u|^2)^2+4|F_H+M_H(\lambda_uv_u+\lambda_dv_d)|^2}\ , \cr
\!\!\!\!\! m_{0_2}^2=M_H^2+\frac{1}{2}(|\lambda_u v_u|+|\lambda_u v_u|^2)^2-\frac{1}{2}\sqrt{(|\lambda_u v_u|^2-|\lambda_u v_u|^2)^2+4|F_H+M_H(\lambda_uv_u+\lambda_dv_d)|^2}\ ,  \cr
\!\!\!\!\! m_{0_3}^2=M_H^2+\frac{1}{2}(|\lambda_u v_u|+|\lambda_u v_u|^2)^2+\frac{1}{2}\sqrt{(|\lambda_u v_u|^2-|\lambda_u v_u|^2)^2+4|F_H-M_H(\lambda_uv_u+\lambda_dv_d)|^2}\ , \cr
\!\!\!\!\! m_{0_4}^2=M_H^2+\frac{1}{2}(|\lambda_u v_u|+|\lambda_u v_u|^2)^2-\frac{1}{2}\sqrt{(|\lambda_u v_u|^2-|\lambda_u v_u|^2)^2+4|F_H+M_H(\lambda_uv_u-\lambda_dv_d)|^2}\ .
\end{eqnarray}

We now define some useful functions. The fermion vacuum polarizations are defined as
\begin{eqnarray}
\Pi_{LL}^{\mu\nu}(M_1,M_2,q^2)=\Pi_{LL}^{\mu\nu}(M_1,M_2,0)+\Pi_{LL}'^{\mu\nu}(M_1,M_2,0)q^2\ , \cr
\Pi_{LR}^{\mu\nu}(M_1,M_2,q^2)=\Pi_{LR}^{\mu\nu}(M_1,M_2,0)+\Pi_{LR}'^{\mu\nu}(M_1,M_2,0)q^2\ ,
\end{eqnarray}
where the prime indicated a derivative with respect to $q^2$ and
\begin{eqnarray}
\Pi_{LL}^{\mu\nu}(M_1,M_2,0)=-2g^{\mu\nu}\left( \frac{1}{16\pi^2}\frac{M_1^4(1-2\ln(M_1^2))-M_2^4(1-2\ln(M_2^2))}{4(M_1^2-M_2^2)}\right)\ , \cr
\Pi_{LR}^{\mu\nu}(M_1,M_2)=2M_1M_2g^{\mu\nu}\left(\frac{1}{16\pi^2}\frac{M_1^2(1-\ln(M_1^2))-M_2^2(1-\ln(M_2^2))}{M_1^2-M_2^2}\right)\ .
\end{eqnarray}
and
\begin{eqnarray}
&&\!\!\!\!\!\!\Pi_{LL}'^{\mu\nu}(M_1,M_2,0)=\frac{g^{\mu\nu}}{8\pi^2}\left( \frac{(9M_1^4M_2^2-3M_1^6)\ln(M_1^2)-9M_2^2M_1^4+M_1^6}{9(M_1^2-M_2^2)^3}+(1\leftrightarrow 2)\right)\ , \cr
&&\!\!\!\!\!\!\Pi_{LR}'^{\mu\nu}(M_1,M_2,0)=2M_1M_2\left(\frac{1}{16\pi^2}\frac{M_1^4-M_2^4 +2M_1^2M_2^2\ln\left(\frac{M_2^2}{M_1^2}\right)}{2(M_1^2-M_2^2)^3}\right)\ .
 \end{eqnarray}
and we have neglected the infinite parts. We also define
\begin{eqnarray}
\Pi_{VV}^{\mu\nu}(M_1,M_2,q^2)=2(\Pi_{LL}^{\mu\nu}(M_1,M_2,q^2)+\Pi_{LR}^{\mu\nu}(M_1,M_2,q^2)) \ .
\end{eqnarray}

The scalar vacuum polarizations depend on
\begin{eqnarray}
&&\!\!\!\!\!\!\!\!\!\!\!\!\!\!\!\!\!\!\!\!\Pi^{\mu\nu}(q^2,m_1,m_2)=
\frac{1}{16\pi^2}\left(\frac{1}{2}(m_1^2+m_2^2)-\frac{m_1^2m_2^2}{m_1^2-m_2^2}\ln\left(\frac{m_1^2}{m_2^2}\right)\right)+F(m_1,m_2)q^2+ \cdots \ .
\end{eqnarray}
where
\begin{eqnarray}
&& F(m_1,m_2)=\frac{1}{3}\frac{1}{16\pi^2}\left(\frac{(6m_1^6-18m_1^4m_2^2)\ln(m_1^2) +5m_2^6+27m_1^4m_2^2}{(m_1^2-m_2^2)^3}+(1 \leftrightarrow 2)
\right)\ .
\end{eqnarray}

The contributions to the vacuum polarizations from the fermions are then
\begin{eqnarray}
&&\Pi_{WW}^{\mu\nu}(q^2)=N_{c'}\frac{g^2}{2}\left(\Pi_{LL}^{\mu\nu}(M_H,m_{\nu_1},q^2) + 2\Pi_{LR}^{\mu\nu}(M_H,m_{\nu_1},q^2)\cos\theta\sin\theta+(1\leftrightarrow 2)\right)\ , \cr
&& \Pi_{ZZ}^{\mu\nu}(q^2)=\frac{g^2N_{c'}}{4\cos^2\theta_W}\sum_{i=1}^2\left[\cos^2\theta\sin^2\theta\Pi_{VV}^{\mu\nu}(m_{\nu_i},m_{\nu_i},q^2) \right. \ ,  \cr
\nonumber && \quad \quad\quad\quad\quad \left. + (\sin^4\theta+\cos^4\theta)\L_{LL}^{\mu\nu}(m_{\nu_i},m_{\nu_i},q^2) +2\sin^2\theta\cos^2\theta\Pi_{LR}^{\mu\nu}(m_{\nu_i},m_{\nu_i},q^2)\right]\ , \cr
&&\Pi_{\gamma\gamma}^{\mu\nu}=(gs_w)^2\Pi_{LL}^{\mu\nu}(M_H,M_H,q^2)\ , \cr
&&\Pi_{\gamma Z}^{\mu\nu}=g^2s_w\left(-\frac{1}{2}+s_w^2\right)\Pi_{VV}^{\mu\nu}(M_H,M_H,q^2)\ .
\end{eqnarray}

The contributions from the scalars are
\begin{eqnarray}
&&\Pi_{WW}^{\mu\nu}(q^2)=\frac{g_2^2}{2}|R_{L_{i1}}^*R_{0_{j2}}-R_{L_{i2}}^*R_{0_{j3}}|^2\Pi^{\mu\nu}(q^2,m_{0_j},m_{e_i})\ , \cr
&&\Pi_{ZZ}^{\mu\nu}(q^2)=\frac{g_2^2}{4\cos^2\theta_W}\left(|R_{0_{i2}}^*R_{0_{j2}}+R_{0_{i3}}^*R_{0_{j3}}|^2\Pi^{\mu\nu}(q^2,m_{0_i},m_{0_j})\right.\cr
&&\quad\quad\quad\quad\quad\quad\quad+|1+2\sin^2\theta_W|^2\Pi^{\mu\nu}(q^2,m_{e_j},m_{e_j})\left. \right)\ , \cr
&&\Pi_{\gamma\gamma}^{\mu\nu}(q^2)=g_2^2\sin^2\theta_W\Pi^{\mu\nu}(q^2,m_{e_i},m_{e_i})\ , \cr
&&\Pi_{\gamma Z}^{\mu\nu}=\frac{g_2^2}{2}\tan\theta_W(1+2\sin^2\theta)\Pi^{\mu\nu}(q^2,m_{e_i},m_{e_i})\ .
\end{eqnarray}

Using these expressions for the vacuum polarizations we can find the $S$ and $T$ parameters from
\begin{eqnarray}
\alpha T=\frac{\Pi_{WW}(0)}{M_W^2}-\frac{\Pi_{ZZ}(0)}{M_Z^2}=\frac{1}{M_Z^2\cos^2\theta_W} \left(\Pi_{WW}(0)-\cos^2\theta_W\Pi_{ZZ}(0)\right)\ ,
\end{eqnarray}
and
\begin{eqnarray}
\alpha S=4s_W^2c_W^2\left(\Pi_{ZZ}'-\frac{c_W^2-s_W^2}{c_Ws_W}\Pi_{Z\gamma}'-\Pi_{\gamma\gamma}'\right)\ .
\end{eqnarray}
The expansions for the $S$ and $T$ parameters can be found in the text.

\section{Determining  Anomalous Dimensions by $a$-Maximization}
\label{sec:a-max}
In this appendix we determine the anomalous dimensions
of the strongly interacting sector by using the so-called $a$-maximization method\,\cite{Intriligator:2003jj} instead of
the Banks-Zaks approximation presented in section\,\ref{sec:conformal} .
The $a$-maximization method states that the conformal $R$ current
appearing in the super-conformal algebra maximizes
a particular 't Hooft anomaly
\begin{equation}
a = \mathrm{Tr}(3R^3 - R).
\end{equation}

In the model of section\,\ref{sec:conformal},
the prescription for conformal $R$-symmetry is given by,
\begin{eqnarray}
R_i= \frac{2}{3}(1 + \gamma_i)\ , \quad (i = 0- 5, H_u)
\end{eqnarray}
which satisfies two conditions,
\begin{eqnarray}
 R_0 + R_L + R_{H_u} = 2\ ,
\end{eqnarray}
and
\begin{eqnarray}
 3 + 3 (R_{\bar D} - 1) + 2 (R_{L} - 1) +  (R_0 - 1) = 0\ .
\end{eqnarray}
Here, we have a priori used $R_{i} = R_{\bar{i}}$ and $R_{H_u} = R_{H_d}$.
Using these constraints, the anomalous dimensions are found by maximizing the $a$-function. The anomalous dimensions found from this procedure are given in section\,\ref{sec:conformal};
\begin{eqnarray}
 \gamma_{0} \simeq -0.18\ , \quad
 \gamma_{L} \simeq -0.23\ , \quad
  \gamma_{\bar D} \simeq -0.29\ , \quad
  \gamma_{H_u} \simeq 0.41\ .
\end{eqnarray}

\section{Renormalization Group Equation  with MSSM Couplings}
\label{sec:RGEapp}
In this appendix, we list the renormalization group equations
used in the analysis in subsection\,\ref{sec:RGE}.
\begin{eqnarray}
\frac{d}{d\ln\mu} g_{3'} &=& \frac{-1}{16\pi^2}\frac{g_{3'}^3}{1-3g_{3'}^2/8\pi^2 }
\left(
3-\frac{g_1^2}{8\pi^2}
-\frac{3g_2^2}{8\pi^2}
-\frac{g_3^2}{\pi^2}
-\frac{2g_{3'}^2}{\pi^2}
+\frac{\l_d^2}{4\pi^2}
+\frac{\l_u^2}{4\pi^2}
\right)\ ,
\nonumber\\
\frac{d}{d\ln\mu} \l_u &=& \frac{\l_u}{16\pi^2}
\left(
6\l_u^2
+ 3 y_t^2
-\frac{3}{5}g_1^2
- 3 g_2^2
-\frac{16}{3}g_{3'}^2
\right)\ ,
\nonumber\\
\frac{d}{d\ln\mu} \l_d &=& \frac{\l_d}{16\pi^2}
\left(
6\l_d^2
-\frac{3}{5}g_1^2
- 3 g_2^2
-\frac{16}{3}g_{3'}^2
\right)\ ,\nonumber\\
\frac{d}{d\ln\mu} y_t &=& \frac{\l_d}{16\pi^2}
\left(
6y_t^2
+3\l_u^2
-\frac{13}{15}g_1^2
- 3 g_2^2
-\frac{16}{3} g_3^2
\right)\ ,
\end{eqnarray}
In our analysis, we have neglected the bottom Yukawa coupling constant.

\end{document}